\documentclass{aa}  
\usepackage{natbib}
\bibpunct{(}{)}{;}{a}{}{,} 
\usepackage{graphicx}
\usepackage{array,tabulary,tabularx,multirow}
\usepackage{txfonts}
\usepackage{lipsum}
\usepackage[english]{babel}
\usepackage{color}
\usepackage[dvipsnames]{xcolor}

\newcommand{\target}[1]{J2102-4145\ }
\newcommand{\Msun}{M$_{\odot}$}
\newcommand{\lppr}{\stackrel{<}{\scriptstyle \sim}}
\newcommand{\lappr}{\raisebox{-0.4ex}{$\lppr$}}
\newcommand{\gppr}{\stackrel{>}{\scriptstyle \sim}}
\newcommand{\gappr}{\raisebox{-0.4ex}{$\gppr$}}
\newcommand{\mesa}{{\sc mesa}}
\newcommand{\bse}{{\sc bse}}
\usepackage{romannum}

\begin{document}

   \title{The double low-mass white dwarf eclipsing binary system J2102-4145 and its possible evolution}


   \author{Larissa Antunes Amaral
          \inst{1,2},  James Munday\inst{3,4}, Maja Vučković\inst{1}, Ingrid Pelisoli\inst{3}, Péter Németh\inst{5,6}, Monica Zorotovic\inst{1}, T. R. Marsh\inst{3}, S.P. Littlefair\inst{7}, V. S. Dhillon\inst{7,8} \and Alex J. Brown\inst{7}
           }

   \institute{Instituto de Física y Astronomía, Universidad de Valparaíso, Gran Bretaña 1111, Playa Ancha, Valparaíso 2360102, Chile\\
   \email{larissa.amaral@postgrado.uv.cl}
              \and
              European Southern Observatory, Alonso de Cordova 3107, Santiago, Chile
              \and
             Department of Physics, University of Warwick, Gibbet Hill Road, Coventry, CV4 7AL, UK
             \and
              Isaac Newton Group of Telescopes, Apartado de Correos 368, E-38700 Santa Cruz de La Palma, Spain  
             \and
             Astronomical Institute of the Czech Academy of Sciences, CZ-251 65, Ond\v{r}ejov, Czech Republic
             \and 
             Astroserver.org, F\H{o} t\'{e}r 1, 8533 Malomsok, Hungary
             \and
             Department of Physics and Astronomy, University of Sheffield, Sheffield, S3 7RH, UK
             \and
             Instituto de Astrofísica de Canarias, E-38205 La Laguna, Tenerife, Spain\\
               }

   \date{Received...; accepted February 15th 2024}

 
  \abstract{In recent years, about 150 low-mass white dwarfs (WDs), typically with masses below $0.4M_{\odot}$, have been discovered. The majority of these low-mass WDs are observed in binary systems as they cannot be formed through single-star evolution within Hubble time. 
  In this work, we present a comprehensive analysis of the double low-mass WD eclipsing binary system J2102-4145.
  Our investigation encompasses an extensive observational campaign, resulting in the acquisition of approximately 28 hours of high-speed photometric data across multiple nights using NTT/ULTRACAM, SOAR/Goodman, and SMARTS-1m telescopes. These observations have provided critical insights into the orbital characteristics of this system, including parameters such as inclination and orbital period. 
  To disentangle the binary components of J2102-4145, we employed the {\sc XT{\fontsize{6}{8}\selectfont
 GRID}} spectral fitting method with GMOS/Gemini-South and X-Shooter data. Additionally, we use the PHOEBE package for light curve analysis on NTT/ULTRACAM high-speed time-series photometry data to constrain the binary star properties.
  Our analysis unveils remarkable similarities between the two components of this binary system. For the primary star, we determine $T_{\text{eff,1}} = 13\,688^{+65}_{-72}$ K, $\log g_1 = 7.36\pm0.01$, $R_1=0.0211\pm0.0002$\,R$_\odot$, and $M_1 = 0.375\pm0.003 M_{\odot}$, while, the secondary star is characterised by $T_{\text{eff,2}} = 12\,952^{+53}_{-66}$ K, $\log g_2 = 7.32\pm0.01$, $R_2=0.0203^{+0.0002}_{-0.0003}$\,R$_\odot$, and $M_2 = 0.314\pm0.003 M_{\odot}$. 
Furthermore, we find a notable discrepancy between $T_{\text{eff}}$ and $R$ of the less massive WD, compared to evolutionary sequences for WDs from the literature, which has significant implications for our understanding of WD evolution. We discuss a potential formation scenario for this system which might explain this discrepancy and explore its future evolution. We predict that this system will merge in $\sim$800\,Myr, evolving into a helium-rich hot subdwarf star and later into a hybrid He/CO WD.}

   
\titlerunning{J2102-4145}
\authorrunning{L. Antunes Amaral et al.} 

\maketitle


\section{Introduction}
White dwarf stars (WDs) are the final remnants of the evolution of all stars with an initial mass below $8.5$-$10 M_{\odot}$, and are thus the end-point of about 95\% of the stars in the Milky Way \citep{Lauffer2018}. 
The observed sample of WDs covers a wide range of mass, ranging from $\sim0.16 M_{\odot}$ to near the Chandrasekhar limit \citep[e.g.,][]{ 2016Tremblay,2017Kepler}. Specifically, for WDs with hydrogen (H) atmosphere (DAs), the mass distribution strongly peaks around $0.6M_{\odot}$, but there is a non-negligible fraction of systems at both the low- and high-mass tails of the distribution \citep[e.g.,][]{2004Liebert,2020Kilic, 2020Tremblay, 2021Kilic}. Many of those low- and high-mass WDs are likely the result of close binary evolution, where mass transfer and mergers commonly occur. 

According to standard stellar evolution theory, the majority of WD stars likely harbour carbon-oxygen (C/O) cores, while the more massive ones can have cores made of oxygen-neon (O/Ne) or neon-oxygen-magnesium (Ne/O/Mg) \citep{Lauffer2018}. However, the population of low-mass WDs, i.e. those with masses below $\sim0.45M_{\odot}$, can harbour either a pure-helium core \citep{2007Panei, 2013Althaus, 2014Istrate, 2016Istrate} or a hybrid core, composed of helium, carbon and oxygen \citep{1985IbenANDTutukov, 2000Han, 2009PradaMoroni, 2019Zenati}.

Low-mass WDs are, as mentioned above, expected to form only through binary interaction since such a low-mass remnant (at least below $M<0.4M_{\odot}$) cannot be formed through single star evolution within Hubble time \citep{1995Marsh, zorotovic2017}.
The formation of those binary systems is believed to occur after an episode of enhanced mass loss in interacting binary systems, before helium is ignited at the tip of the red giant branch (RGB, \citealt{2013Althaus, 2016Istrate, 2019Li}). This interaction will leave behind a naked He remnant, which will later become a He WD, or a hybrid He-C/O WD if the mass is sufficient to ignite He after shedding the envelope. In the latter scenario, the star goes through a hot subdwarf phase before becoming a WD \citep{2019Zenati}. The binary evolution scenario is currently supported by observations since most low-mass WDs are observed to be in binary systems \citep{2016Brown_ELM7}. The fraction of them that seems to be single is consistent with merger scenarios in close binary systems \citep{zorotovic2017}.

The loss of the envelope in the progenitors of low-mass WDs can occur due to either (i) common-envelope evolution \citep[CE,][]{1976Paczynski, 1993IbenandLivio} or (ii) a stable Roche-lobe overflow episode \citep[RLOF,][]{1986Iben}. Typically, binary systems with short orbital periods (hours to a few days) arise predominantly from CE interactions, whereas those with longer orbital periods are more prone to form through stable RLOF. 

Extremely-low mass (ELM) WDs are He-core WDs with a mass below $\sim0.3M_{\odot}$, low enough to ensure that helium was never ignited \citep{2010Brown_ELM1}. There are currently more than 150 known ELMs WDs \citep[e.g.,][and references therein]{2022Brown_ELM9,2023ELMSouth2}, as well as ELM WD candidates \citep{2019PelisoliandVos,Wang2022}.

Low-mass and ELM WDs display a variety of photometric variations, including Doppler beaming \citep{2010Shporer, 2014Hermes}, tidal distortions, pulsations \citep{2012Hermes}, and eclipses {\citep{ 2010Steinfadt, 2015Gianninas_ELM6}. 
Low-mass WDs in binary systems provide a unique opportunity to significantly enlarge the known population of merging WDs in the Galaxy, representing a collection of objects that contribute to a steady foreground of gravitational waves. Identifying and characterising compact binary systems in the Milky Way will help characterise the noise floor that may otherwise impede on LISA’s ability to detect gravitational waves \citep{2017Amaro-Seoane_LISA}. Moreover, short-period binary WDs are also potential progenitors of Type \Romannum{1}a supernova \citep{1984Webbink, 1984Iben_Tutukov, 2007Bildsten}. The population of low-mass WDs in eclipsing systems are a gold standard in astrophysics, allowing for the most precise measurements of the stellar and binary parameters, with precision that can reach below the per cent level \citep{2023Brown}.

Around a dozen low-mass pulsating WDs have been discovered in the last decade \citep{2012Hermes, 2013Hermesa, 2013Hermesb, 2015Kilic, 2015Bell, 2017Bell, 2018Pelisolib,  2020Parsons, 2021Guidry, 2021Lopez}, most of which correspond to ELM WDs (also known as ELMVs due to their variability). Pulsations have greatly sparked the interest in low-mass WDs, as it provides a unique opportunity to explore the internal structure of WDs at cooler temperatures (6.0 $\lesssim$ $\log{g}$ $\lesssim$ 6.8 and 7800~K  $\lesssim$ $T_{\text{eff}}$ $\lesssim$ 10\,000~K) and much lower masses ($< $0.4$M_{\odot}$), than common pulsating DA WDs (ZZ Ceti). Likewise, the detection of a pulsating WD within an eclipsing binary system can be a potent reference point for empirically constraining the core compositions of low-mass stellar remnants. 

So far, there have been only three pulsating WDs found that belong to a detached eclipsing binary system: the first pulsating low-mass WD ($\sim0.325\,M_{\odot}$) in a double-degenerate eclipsing system \citep{2020Parsons} and the first two ZZ Ceti WDs found in eclipsing WD + main sequence post-common envelope binaries \citep{2023Brown}.
Those systems are powerful benchmarks to constrain empirically the core composition of low-mass stellar remnants and investigate the effects of close binary evolution on the internal structure of WDs. 

Very recently, in the ELM Survey South II, \citet{2023ELMSouth2} have reported the discovery of another eclipsing double WD binary J210220.39-414500.77 (hereafter J2102-4145), which most likely contains two low-mass WDs, similar to the system from \citet{2020Parsons}. The authors have established this system as a double-lined, double-degenerate, eclipsing WD binary with well-constrained radial velocity (RV) measurements and orbital period. \citet{2023ELMSouth2} derived atmospheric and orbital parameters as well as masses of each star. While one can estimate the model-dependent radius of each star using \citet{2023ELMSouth2} results they have not provided the direct radius measurements.  
Also, it is worth noting that \citet{2023ELMSouth2} did not use the eclipses visible in the TESS photometry to derive individual star radii, as TESS data is compromised by contamination from a nearby brighter source due to TESS pixel scale of 21’’ per pixel.}. In this work, we present an in-depth analysis of J2102-4145, including new spectroscopic and photometric data, which allowed us to constrain the orbital parameters and confirm that both WDs in the system have a mass below 0.4$M_{\odot}$. We explored potential insights for a pulsation period and found no conclusive evidence supporting pulsations in either of the two WDs.

\section{Observations}

\subsection{High speed photometry}
High-speed time-series photometry was obtained using three different instruments: Goodman at the 4.1~m Southern Astrophysical Research (SOAR) Telescope in May 2021, ULTRACAM at the 3.5~m ESO New Technology Telescope (NTT) in July 2021, and the Apogee F42 camera at the  1~m Small and Moderate Aperture Research Telescope System (SMARTS) in May 2022. With Goodman/SOAR we used the Blue Camera with the S8612 red-blocking filter. We used read-out mode 200 Hz ATTN2 with the CCD binned 2$\times$2 and integration time of 5 seconds. For ULTRACAM/NTT \citep{Dhillon2007}, dichroic beam splitters allow simultaneous observations in three different filters. We observed using the Super SDSS filters $u_{s}$, $g_{s}$, $i_{s}$, and $u_{s}$, $g_{s}$, $r_{s}$, for which the $u_{s}$ band had 3 times the exposure time of the $g_{s}$, $r_{s}$ and $i_{s}$ bands (9~s and 3~s respectively, with 24~ms dead time between each exposure) (see Table \ref{table:phot_log}). Lastly, we observed with SMARTS-1m with 1$\times$1 binning, SDSSg filter with an exposure time of 25~s. Details of each of our runs can be found in Table \ref{table:phot_log}.

The SOAR and SMARTS-1m photometric data were reduced using the \textsc{IRAF}\footnote{https://iraf-community.github.io/} software, with the \textsc{DAOPHOT} package to perform aperture photometry. All images were bias-subtracted, and flat field corrected using dome flats. Neighbouring non-variable stars of similar brightness were used as comparison stars (Gaia EDR3\,6581252815151045248 and Gaia EDR3\,6581249104299291648 for SOAR and SMARTS data reduction, respectively) to perform the differential photometry. We then divided the light curve of the target star by the light curves of all comparison stars to minimise the effects of sky and transparency fluctuations. 

All ULTRACAM data was reduced using the HiPERCAM reduction pipeline \citet{2021Dhillon}. Bias, flat and dark frames were taken and applied to the science images. 
A target's photometry was extracted via standard differential aperture photometry. A variable aperture radius equal to $1.8\times$ the centroid full-width at half-maximum was applied. The non-variable comparison star Gaia DR3 source ID 6581252815151045248 was used for all filters.

Using the ULTRACAM data and optimising the ephemeris through eclipse timing (see Section \ref{subsec:PHOEBEmethod}), we constrain the orbital ephemeris, centred on the primary mid-eclipse, to be:
\begin{equation}
    \text{BMJD}_{\text{min}} = 59410.360987(30)+ 0.1002087525(10) E  
\end{equation}

\begin{table*}[h!]
\caption{Journal of photometric observations from ground-based facilities. We list the telescope/instrument, start date of observation, total observation time, cadence time, exposure time and filter used in columns 1, 2, 3, 4, 5 and 6, respectively. Note that for ULTRACAM the cadence and exposure time are equal since we used frame-transfer CCDs, giving only a 24\,ms dead time.}             
\label{table:phot_log}      
\centering          
\begin{tabular}{c c c c c c}     
\hline     
Telescope/Instrument & Start(BJD) & Duration(h) & Cadence(sec) & Exposure time (sec) & Filter\\ 
\hline   
SOAR/GoodMan &2459351.759849080 
& 4.32 & 10 & 5 & S8612\\
SMARTS-1m &2459725.751409210 
& 3.6 & 30 & 25& SDSSg \\
NTT/ULTRACAM & 2459410.857675513
& 12.4 & 9/3/3 & 9/3/3 & $u_{s} / g_{s} / i_{s}$\\

NTT/ULTRACAM &  2459841.513171684 
& 7.5 & 9/3/3 & 9/3/3 & $u_{s} / g_{s} / r_{s}$\\

\hline 
\end{tabular}
\end{table*}


\subsection{Spectroscopy}
\label{SUBSection_spectroscopy}
We obtained time-resolved spectroscopy for \target\ using two different telescopes and instruments: the GMOS spectrograph \citep{Hook2004, GMOS-South} on Gemini South 8.1 m telescope in queue mode, and the X-Shooter at the 8.2 m Very Large Telescope (VLT) \citep{2011Vernet-xshooter}. 
For the GMOS spectrograph, a total of 18 exposures were taken with a 1$\arcsec$ slit. We binned the CCD by a factor of two in both dimensions and used a 600 l/mm grating. The amplifier number 5 on GMOS has been showing abnormally high counts, plus noise structures on the rest of the CCD2, since March 2022. Our data has been partially affected by this issue. Thus, to dislocate the position of the two gaps between the CCDs in GMOS and to compensate for the gap in amplifier number 5, exposures centred at 490nm, 520nm and 550nm were taken (coverage 345-647nm, 373-677nm and 393.4-707nm, respectively). These configurations resulted in a resolving power R $\sim$ 844.
For all exposures, arc-lamp and flat exposures were taken before and/or after each science exposure to verify the stability. For the wavelength calibration, a CuAr lamp was taken after each round of exposures, at the same telescope position as the science frames. We obtained 250~s long exposures under IQAny or IQ85 percentile conditions with GMOS, resulting in signal-to-noise ratio (S/N) of about 7–30 per exposure.

The Gemini spectroscopic data was reduced using the \textsc{Gemini-GMOS IRAF} package. This included the subtraction of an averaged bias frame and division by a flat field that was previously normalised by fitting a cubic spline of high order to the response function. A two-dimensional wavelength calibration solution was obtained with respect to a CuAr lamp. After extraction, the spectra were corrected for the instrument response function using the corresponding data of a spectrophotometric standard. Since only one spectrophotometric standard was observed per semester on Gemini-South, we did not correct it by the zero-point. In other words, the flux calibration is only relative and not absolute.

The X-Shooter observations were carried out using a 1$\arcsec$ slit for the UVB arm and a 0.9$\arcsec$ slit for the VIS and NIR arms. A total of 70 consecutive exposures were taken, in which 25 exposures of 288\,s were taken with the UVB arm (300-559.5 nm), 25 exposures of 300\,s with the VIS arm (559.5-1024 nm), and 18 of 480\,s with the NIR arm (1024-2480 nm). The UVB and VIS arms were binned by 2$\times$2, while the NIR arm was not binned. Given that the seeing was much better (on average 0.4$\arcsec$) than the slit width, we have calculated the resolution directly from the spectra which at H$\alpha$ is about 0.4 $\AA$ for VIS arm. 
The X-shooter data was automatically reduced with an ESO pipeline\footnote{http://www.eso.org/sci/software/reflex/}.

Both sets of observations (GMOS and X-Shooter) were performed covering slightly more than one full orbit of the binary. For GMOS/Gemini-South the observations were carried in two separate nights (July 7th and 9th 2022), while X-Shooter observations were carried out in one night (August 8th 2022).

\section{Methods}
\subsection{ \textsl{XT{\fontsize{8}{8}\selectfont
 GRID}}: Spectral analysis}
\label{subsec:Spectralfit_method}
We applied the steepest-descent iterative fitting procedure {\sc XT{\fontsize{7}{8}\selectfont
 GRID}} \citep{2012nemeth} to disentangle the binary components of \target\ from both the Gemini/GMOS and X-Shooter spectra. 
To reproduce the binary components, we used interpolated models from the $\langle$3D$\rangle$ pure-hydrogen non-Local Thermodynamic Equilibrium DA WD grid of \cite{2013tremblay, tremblay15} with line profiles from \citet{tremblay2009}. 
Those models are suitable for both members of the system. 

Two parallel threads of the fitting procedure were started, each modelling one component of the binary and including the actual model for the secondary from the other thread. 
{\sc XT{\fontsize{7}{8}\selectfont
 GRID}} is designed to adjust on both threads the $T_{\rm eff}$, $\log{g}$, abundances, projected rotation, and the flux contributions to the composite spectrum. 
The observed spectra were fitted with a linear combination of the two components suitably shifted in RV space. 
The eclipse depths (see Figure \ref{LC_eclipses}) imply that the flux contributions of the members must be very similar, which is a major challenge in wavelength space decomposition.
The nearly indistinguishable components make RV measurements difficult from the broad lines and low-resolution spectra belonging to GMOS/Gemini spectra. 
Thus, due to the low spectral resolution, we neglected the projected rotation and the pure H models simplified the analysis further for all GMOS data. 
However, the same does not apply to X-Shooter spectra. 
Being a medium-resolution echelle spectrograph, the X-Shooter data set has been prioritized in the spectral analysis. 
The higher resolving power of X-Shooter is a key to resolving the line cores and determining the RV.


We started with a fixed 50\% contribution from each star and changed the flux contribution according to the relative luminosities of the stars.  
For an ideal solution, the two threads must converge the same results within their respective error bars. 

\begin{figure*}[h!]
   \centering
   \includegraphics[width=\textwidth]{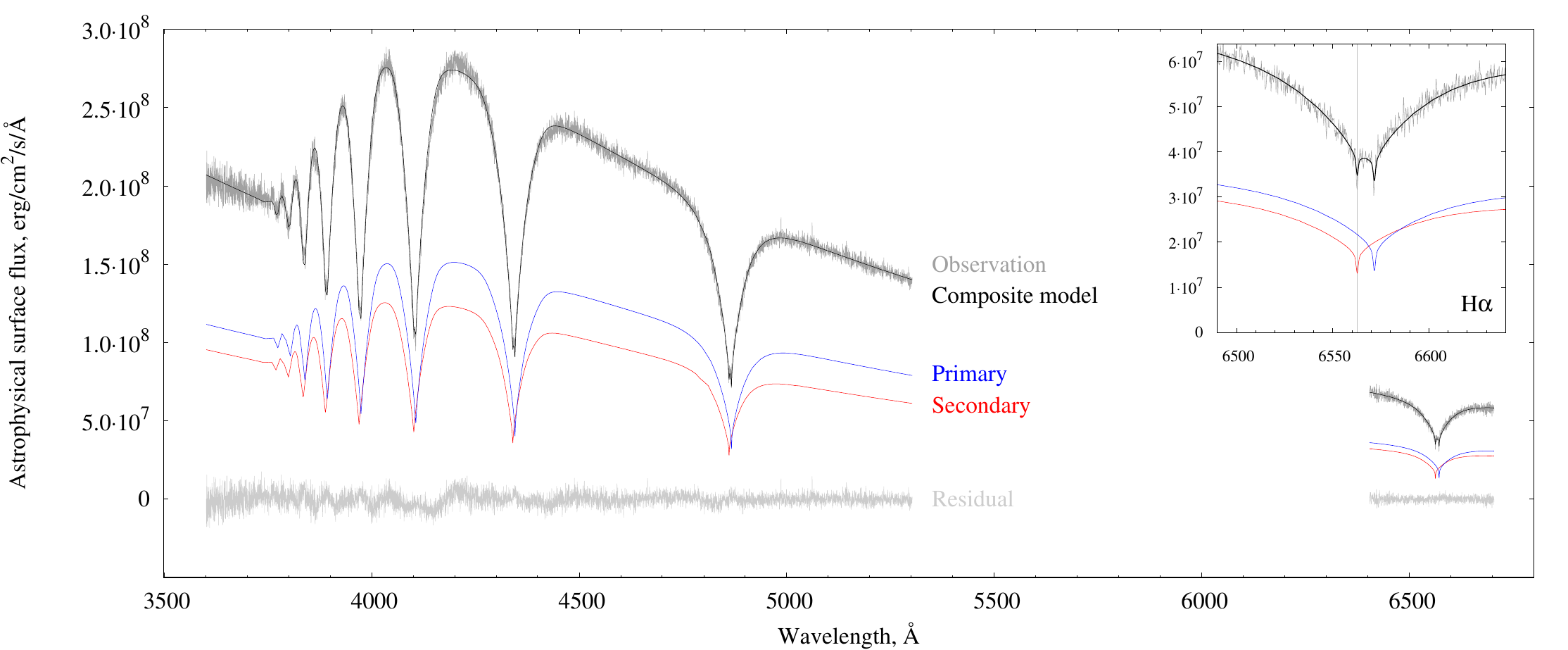}
      \caption{
      Best-fit {\sc XT{\fontsize{6}{8}\selectfont
 GRID}}/WD model to the VLT/X-Shooter spectrum of \target\ at orbital phase $\phi=0.25$ (maximum radial velocity difference) in the rest frame of the secondary. 
      The observed spectrum can be reproduced by two nearly identical WD components ({\sl red and blue lines}). 
      At $\phi=0.25$ and $0.75$, the components show the largest radial velocity difference and the double absorption core of the H$\alpha$ reveals the binarity ({\sl inset}). 
      The phase-resolved spectral coverage for the entire orbit is available in Figure\,\ref{H_alpha_beta_xshooter}. 
          }
\label{SpFit_xshooter}
\end{figure*}

Once the spectral analysis converges and the relative changes of all stellar and binary parameters decrease below 0.5\%, {\sc XT{\fontsize{7}{8}\selectfont
 GRID}} calculates the errors by changing each parameter in one dimension until the chi-square change corresponds to the 60\% confidence limit. 
Figure\,\ref{SpFit_xshooter} shows the best-fit composite model for a representative VLT/X-Shooter spectrum at orbital phase $\phi=0.25$.  
The inset plot in Figure\,\ref{SpFit_xshooter} depicts the RV differences for the primary and secondary components, revealing the binary nature. The same plot, but for the Gemini-GMOS data, can be seen in Figure \ref{SpFit_gemini} in the appendix.

\begin{figure*}[h!]
\centering
   \includegraphics[width=\textwidth]{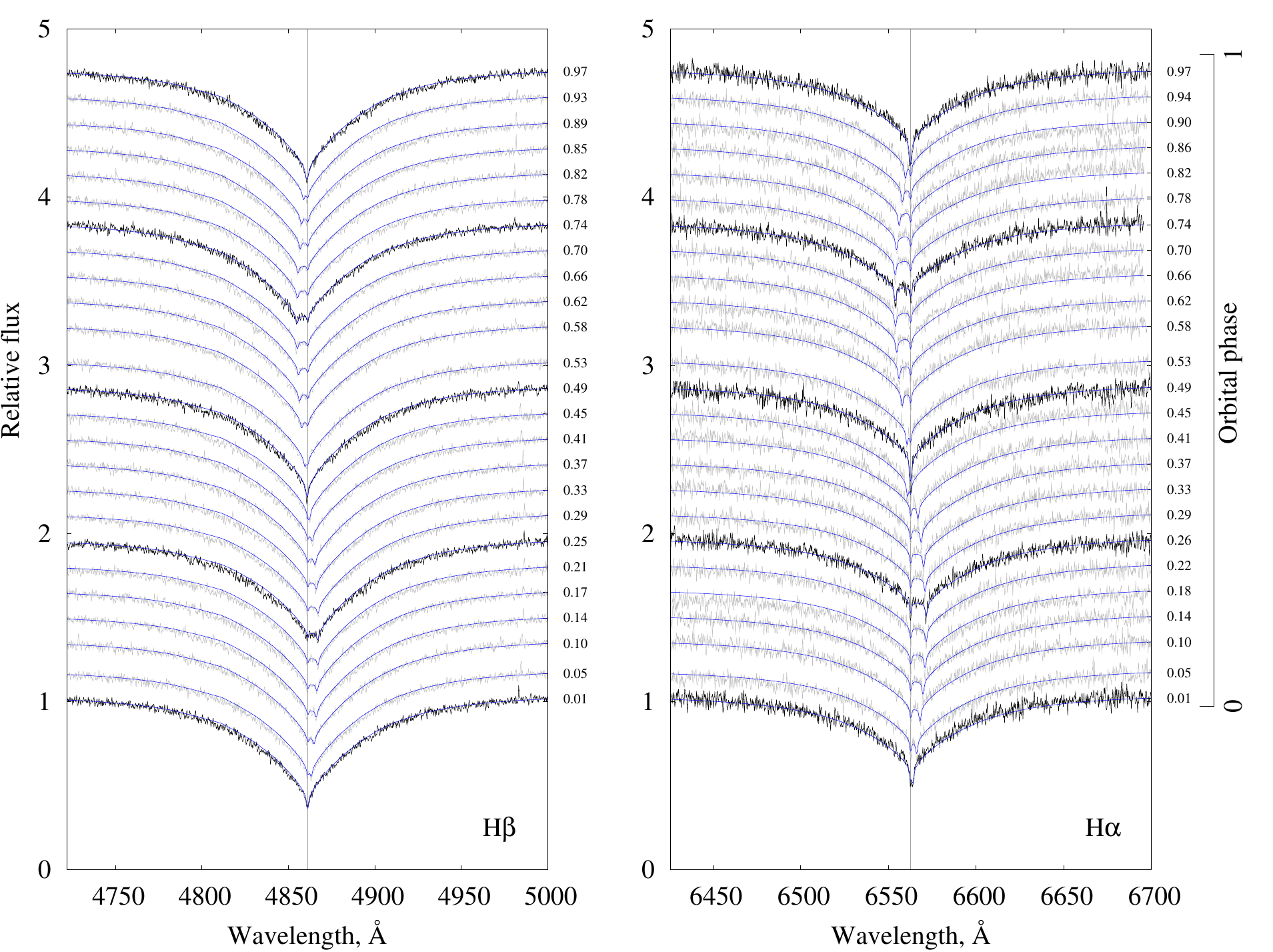}
     \caption{
     Best-fit {\sc XT{\fontsize{6}{8}\selectfont
 GRID}}/WD models and the spectral evolution of the composite H$\alpha$ (right panel) and H$\beta$ (left panel) composite line profiles in the VLT/X-Shooter observations. 
     Observations are in grey and black, where the black colour distinguishes orbital phases nearest to the conjunctions and quadratures.
     The composite model is in blue. 
     We show here the continuum normalised spectra for easier stacking. 
     All fits have been done on the original flux-calibrated data.}
     \label{H_alpha_beta_xshooter}
\end{figure*}

The best fits for all X-Shooter spectra can be seen in Figure\,\ref{H_alpha_beta_xshooter}. 
The same plot, but for GMOS data is shown in Figure \ref{H_alpha_beta_GMOS}, in the appendix. 
The $H_{\alpha}$ and $H_{\beta}$ composite line profiles are shown in the left and right panels, respectively, and are organised by the orbital phase.
The final surface parameters and their errors are listed in Table\,\ref{tab:SpPar}.  

\begin{table}
    \centering
    \caption{Atmospheric parameters from the X-Shooter spectra. }
    \begin{tabular}{lcc}
    \hline
    Parameter                  &   Primary      &  Secondary \\
    \hline
    $T_{\rm eff}$ (K)          & 13790$\pm$670  & 12710$\pm$180\\
    $\log{g}$ (cm\,s$^{-2}$)   & 7.25$\pm$0.19& 7.21$\pm$0.08\\ 
    \hline
    \end{tabular}
    \label{tab:SpPar}
\end{table}


\subsection{PHOEBE: Light curve fitting method}
\label{subsec:PHOEBEmethod}
Since it has the highest signal-to-noise ratio of our full span of photometric data sets and consists of fast-cadence observations, we modelled the ULTRACAM light curves using the PHysics Of Eclipsing BinariEs v2.4 \citep[PHOEBE,][]{Prsa2016PHOEBE2_0, Horvat2018PHOEBE2_1, Jones2020PHOEBE2_2, Conroy2020PHOEBE2_3} package to constrain all components of the binary. In addition, we simultaneously modelled the multi-band photometry and the primary and secondary RVs. 

We assign both stars to have blackbody atmospheres and allow the secondary albedo to vary in each passband. Interpolated limb darkening coefficients following a power law were assigned to each star from the 3-dimensional WD models of \citet{Claret3Dcoefficients2020andDoppler}. Interpolated gravity darkening coefficients were assigned using the tables of \citet{Claret1Dcoefficients2020}. The PHOEBE~2.2 Doppler beaming function was reinstated for our study (see \citealt{2023Munday}) and interpolated Doppler beaming coefficients from \citet{Claret3Dcoefficients2020andDoppler} were passed. For all of these interpolated components, the coefficients are unique for each of the Super SDSS filters and the respective tables were used for filters $u_s$, $g_s$, $r_s$ and $i_s$. The impact of smearing across a finite exposure time was corrected for assuming 9\,s, 3\,s, 3\,s and 3\,s for the $u_s$, $g_s$, $r_s$ and $i_s$, respectively. Light-travel-time effects were handled with the PHOEBE 
code, accounting for R{\o}emer delay. The synthetic RVs incorporate the impact of the stars' surface gravity on the measured RVs and are determined from the centre of light of the system at a given phase. 

In search for a system solution, we implemented an MCMC algorithm using the Python package \textsc{emcee} \citep{emcee2013}, where a series of walkers converged through minimisation of the $\chi^2$ between the observed light curves and the synthetic (and scaled) model light curves. The post burn-in posterior distribution of the MCMC was analysed using the \textsc{corner} python package \citep{corner} and the resultant plot can be seen in Figure \ref{corner_plot} in the appendix. The free parameters that we employed were combinations of the effective temperature ($T_{\text{eff}}$), the radii and the masses of both stars, the secondary albedo in each passband, the systemic velocity and the system inclination. 

The primary albedo was fixed at 1.0 in all bands as allowing both the primary and secondary to vary would lead to degeneracy between parameters and perhaps a nonphysical local minimum. There is also degeneracy between the star temperatures and the secondary albedo. To alleviate this, we first modelled with a fixed primary temperature as obtained from spectroscopy (see Table \ref{tab:SpPar}) and found secondary albedos of $1.04\pm 0.03$, $1.01\pm 0.02$, $1.00\pm 0.02$ and $1.03\pm0.05$, for $u_s$, $g_s$, $r_s$, $i_s$ filters, respectively. To obtain a final model, we allowed the primary effective temperature to vary freely and forced Gaussian priors with the aforementioned secondary albedos. In this setup, the primary temperature is largely constrained by the interpolation of limb/gravity darkening models and its effect on the light curve morphology, while the secondary temperature is largely dependent on these same parameters and the relative eclipse depth.

Furthermore, the eccentricity of the orbit was found to be very low, with an upper bound of $e < 5\times10^{-5}$ (2$\sigma$) or $2\times10^{-5}$ (1$\sigma$), suggesting a nearly circular orbit.

\begin{figure*}
\centering
   \includegraphics[width=17cm]{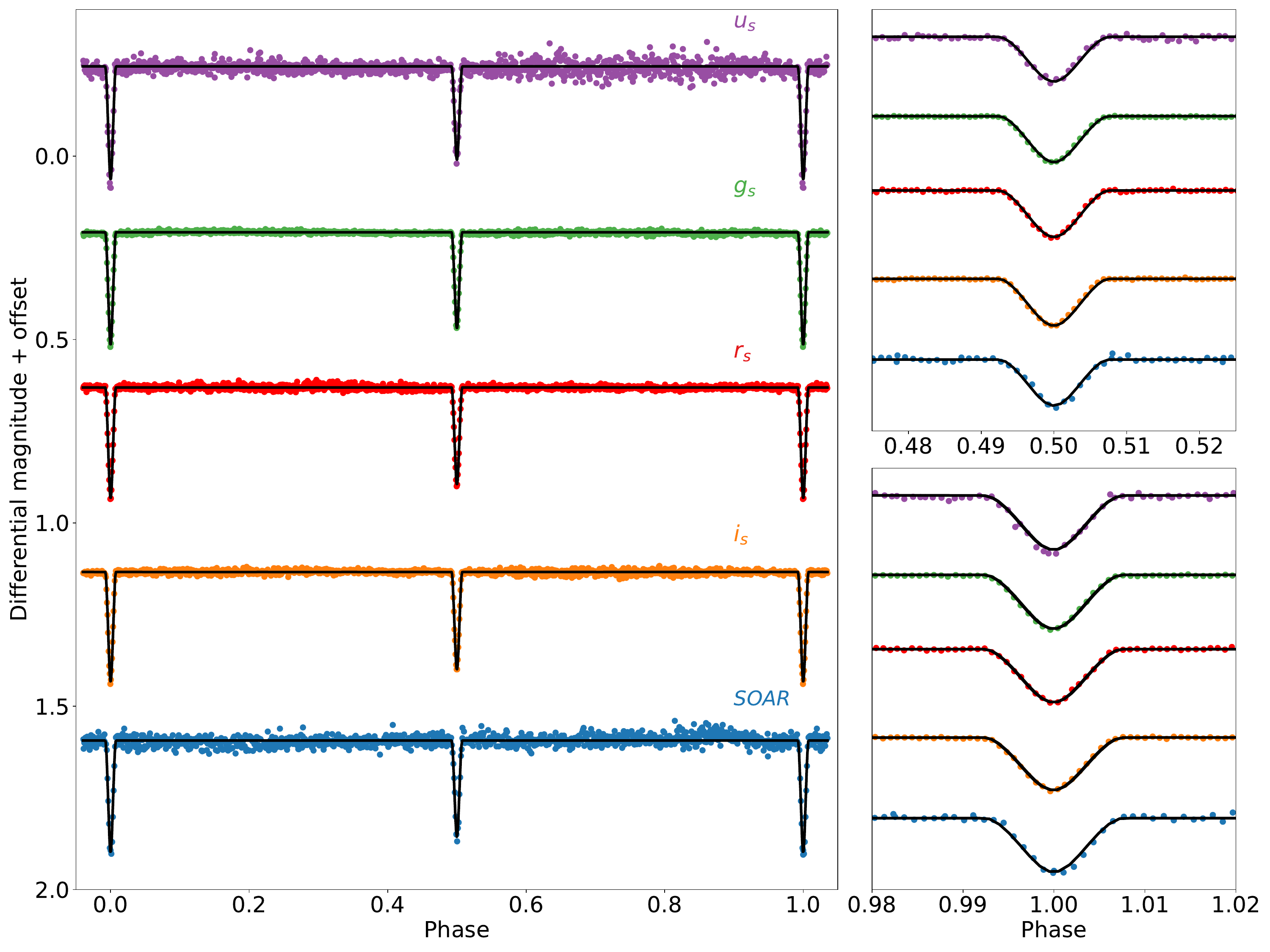}
     \caption{ULTRACAM $u_{s}$, $g_{s}$, $r_{s}$ $i_{s}$, and SOAR phased-folded light curves (coloured points) of \target\  with the best-fit light curve model over-plotted in black (left panel). Zoomed-in primary and secondary eclipses are also shown in the bottom-right and top-right panels, respectively.}
     \label{LC_eclipses}
\end{figure*}

Our resultant best-fit parameters are given in Table~\ref{tab:Binary_params}. In addition, the ULTRACAM $u_s$, $g_s$, $r_s$, $i_s$ and SOAR phase-folded light curves are shown in Figure \ref{LC_eclipses} with the best-fit for each light curve shown as a black line. We also depict the primary and secondary eclipses zoomed-in in the same Figure. It is important to note that the SOAR and SMARTS-1m light curve were not used for this fit due to its lower cadence ($\sim$30\,s), resulting in a limited number of data points during each eclipse also a high smearing effect. 
\renewcommand{\arraystretch}{1.1}
\begin{table}
    \centering
    \caption{\target\ system parameters from light curve fitting to the ULTRACAM photometry. Errors are quoted as the 16$^{\text{th}}$ and 84$^{\text{th}}$ percentiles of the post-burnin MCMC posteriors. It's worth mentioning that the reported errors on these parameters do not include systematic errors, such as period error, black body modelling, and flux calibration. The results from the corner plot should be interpreted with due consideration.}
    \begin{tabular}{l r}
    \hline
    Parameter (Photometry) & Value \\
    \hline
        Period & 0.1002087525(10)\,d\\
         Primary mass  & $M_1=0.375 \pm0.003\,\text{M}_\odot$\\
         Primary temperature & $T_1=13\,688^{+65}_{-72} $\,K\\
         Primary radius  & $R_1=0.0211\pm0.0002$\,R$_\odot$\\
         Secondary mass  & $M_2=0.314\pm0.003$\,\text{M}$_\odot$\\
         Secondary temperature & $T_2=12\,952^{+53}_{-66}$\,K\\
         Secondary radius  & $R_2=0.0203^{+0.0002}_{-0.0003}$\,R$_\odot$\\
         Inclination  & $i=88.693\substack{+0.006\\ -0.005}$\,deg\\
         Primary surface gravity & $\log g_1=7.36\pm0.01$\,dex\\
         Secondary surface gravity & $\log g_2 = 7.32\pm0.01$\,dex\\
         \hline
    \end{tabular}
    \label{tab:Binary_params}
\end{table}


\subsection{Light curve analysis: Looking for periodicities}
\label{subsec:lc_analysis}
To look for periodicities in the light curves due to pulsations, we used two different methods in order to remove the eclipses. The first consisted of solely masking the eclipses given that the duration of each eclipse is only 2\,min. The second was subtracting the binary models from PHOEBE, and then using the residual light curve. In both cases, after each process, we calculate the Fourier Transform (FT), using the software Period04 \citep{2004Lenz_Breger}.  We accepted a frequency peak as significant if its amplitude exceeds an adopted significance threshold. In this work, the detection limit (dashed line in Figure \ref{LC_pulsation}, \ref{LC_pulsation_masked} and \ref{LC_pulsation_masked_noDoppler} ) corresponds to the 1/1000 False Alarm Probability (FAP), where any peak with amplitude above this value has 0.1\% probability of being a false detection due to noise. The FAP is calculated by shuffling the fluxes in the light curve while keeping the same time sampling, and computing the FT of the randomised data. This procedure is repeated N/2 times, where N is the number of points in the light curve. For each run, we compute the maximum amplitude of the FT. From the distribution of maxima, we take the $0.999$ percentile as the detection limit \citep{2020Romero_hotdav}. The internal uncertainties in frequency and amplitude were computed using a Monte Carlo method with 100 simulations with \texttt{PERIOD04}, while uncertainties in the periods were obtained through error propagation. 

To ensure that the observed signal in our masked method is not influenced by the Doppler beaming effect, we generated a new PHOEBE model for SOAR and ULTRACAM data in $u_{s}$, $g_{s}$, $i_{s}$, and $r_{s}$ bands removing the Doppler effect contribution. We thus subtract this new binary model from PHOEBE and subsequently analyse the residual light curve. This approach mirrors our previous methodology. This adjustment aimed to investigate and eliminate the possibility that the signals observed with masked eclipses (see Figure \ref{LC_pulsation}) were not caused by the Doppler beaming effect.

From the masking method, we found significant peaks in most of our data sets (see Figure \ref{LC_pulsation}). Yet, the same peaks were not seen in the FT for the residual light curves obtained by subtracting the binary model. This holds true for both cases, whether the Doppler beaming effect was taken into account during the PHOEBE model calculation (Figure \ref{LC_pulsation_masked}) or was not considered (Figure \ref{LC_pulsation_masked_noDoppler}).


\section{Results and Discussion}
\target\ is an eclipsing double-lined and double-degenerate WD binary, as confirmed by \citet{2023ELMSouth2}. This star has a magnitude of $G=15.7$ and is located at a Gaia-derived distance of $163.4 \pm 1$ pc \citep{2021Bailer-Jones}. 
Our investigation encompassed an extensive observational campaign, resulting in the acquisition of approximately 28 hours of high-speed photometric data across multiple nights conducted using three different telescopes. These observations have provided critical insights into the orbital characteristics of this system, including parameters such as inclination and the orbital period, which was previously known from \citet{2023ELMSouth2}. Notably, our study yields highly precise parameter measurements, with a particular focus on the masses. Furthermore, these observations allowed us to determine the radii of both stellar components while simultaneously constraining the mass of each star.

Although \citet{2023ELMSouth2} provided a comprehensive spectral analysis of this system, in this study we present an additional spectral analysis, by using our time-series spectroscopy data acquired independently with two different telescopes Gemini South/GMOS and VLT/X-Shooter. This supplementary spectral analysis not only enhances the accuracy of the derived system's fundamental properties but also enriches our understanding through comparative analysis.

\subsection{The masses, radii and effective temperatures of both WDs}
\label{subsec:results_fit}
\begin{figure}
\centering
   \includegraphics[width=8.8cm]{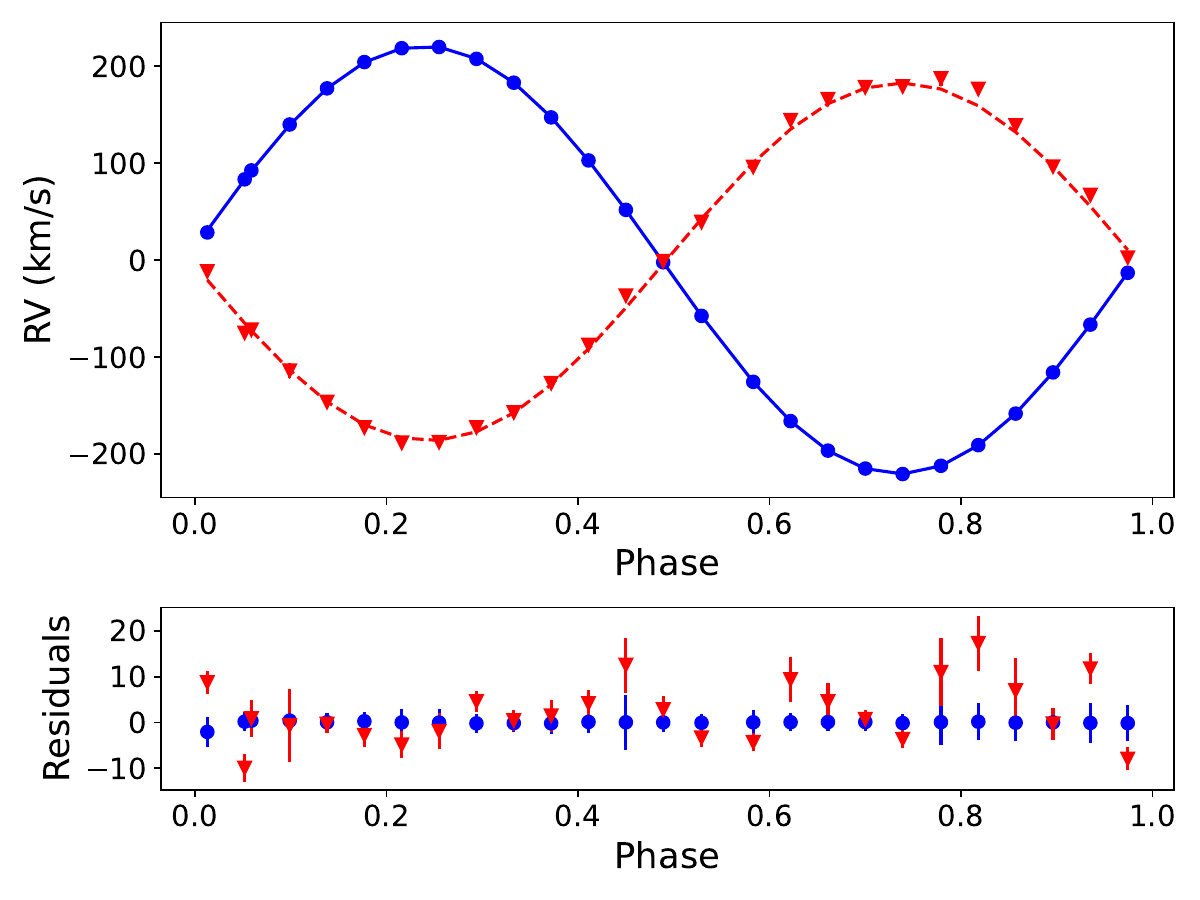}
     \caption{Orbital solutions for each component of the double-lined binary \target\ using X-Shooter VIS arm data. The blue circle points represent the RV for the primary (hottest and more massive) star, while the red triangles represent the RVs for the secondary (cooler) star. }
     \label{rv_xshooter}
\end{figure}

We applied the iterative fitting procedure to disentangle the binary components from the Gemini/GMOS and X-Shooter spectra, in conjunction with the model of the multi-band photometry ULTRACAM light curves. We determined the $T_{\text{eff}}$ and $\log g$ for both stars, which yielded the following parameters: $T_{\text{eff,1}} = 13\,688^{+65}_{-72}$ K, $\log g_1 = 7.36\pm0.01$ for the primary star, and $T_{\text{eff,2}} = 12\,952^{+53}_{-66}$ K, $\log g_2 = 7.32\pm0.01$ for the secondary star. As well as the masses $M_1 = 0.375\pm0.003 M_{\odot}$ for the primary star and $M_2 = 0.314\pm0.003 M_{\odot}$ for the secondary star.
Our mass estimates and $T_{\text{eff}}$ values are consistent with previous measurements, such as those reported by \citet{2023ELMSouth2}.
Furthermore, our light curve fitting method enabled us to measure the radii of each individual WD. We obtained the following radii: $R_1=0.0211\pm0.0002$\,R$_\odot$ and $R_2 = 0.0203^{+0.0002}_{-0.0003},R_{\odot}$. Additionally, we derived RVs from the X-Shooter spectrum using only VIS arm data set. Those results can be seen in Figure \ref{rv_xshooter} and Table \ref{tab:RV_values}, where the RV values for the primary star ($RV_1$) are depicted as blue circles and red triangles for the secondary ($RV_2$). We also show their respective sinusoidal fit, which yields to semi-amplitude values of $K_1 = 220.8\pm0.7$ km/s and $K_2 = 184.6\pm0.8$ km/s. Our RV semi-amplitude values are comparable with the ones found by \citet{2023ELMSouth2}. The bottom plot of Figure \ref{rv_xshooter} shows the residuals for each RV curve. Those results are also consistent with previous measurements done by \citet{2023ELMSouth2}.

\begin{figure}
\centering
   \includegraphics[width=8.8cm]{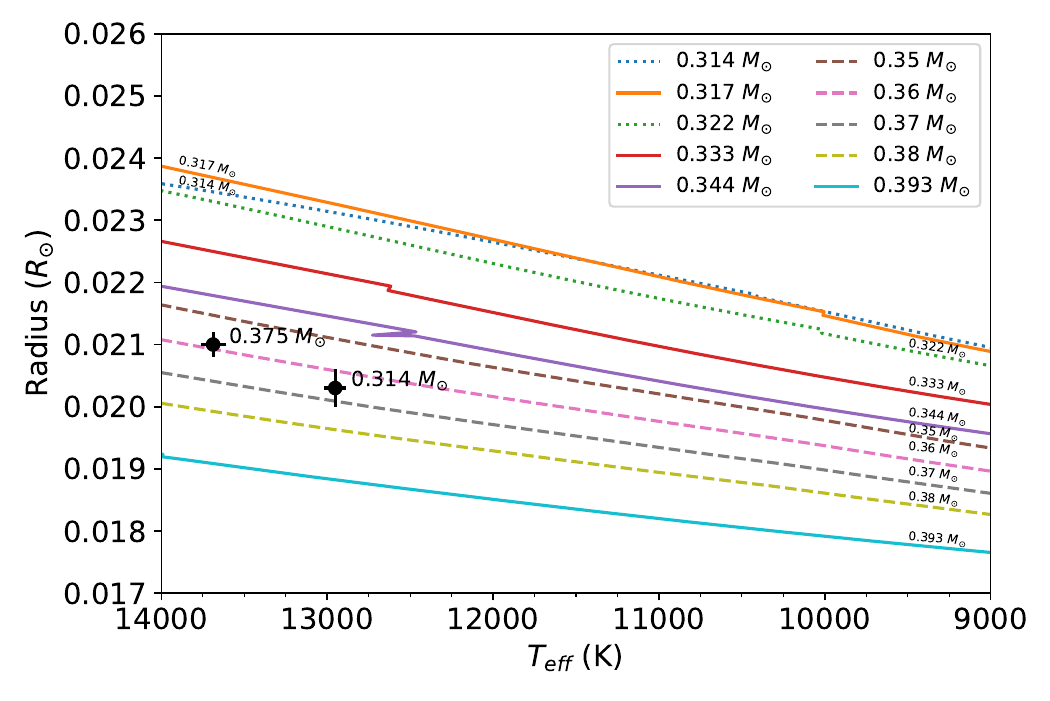}
     \caption{Stellar radius as a function of the effective temperature for He-core WD evolutionary sequences with stellar masses going from 0.317 to 0.393$M_{\odot}$. The solid lines represent evolutionary models from \citet{2016Istrate} for metallicity of Z=0.02, dotted lines are \citet{2016Istrate} models for metallicity Z=0.001 and the dashed lines are models from \citet{2013Althaus} for metallicity Z=0.001. Both stars of the eclipsing system \target\, are shown in black circles, with their respective radius and $T_{\text{eff}}$ values found in this work.}
     \label{teff_radius}
\end{figure}

A notable discrepancy is observed between the temperature and radius of the less massive WD, which appeared to be too hot for its derived mass and radius. This can be seen once we analysed the radius-$T_{\text{eff}}$ diagram in Figure \ref{teff_radius} in which the more massive star falls within the evolutionary models for the same derived mass, within its uncertainties. However, the less massive star should be much colder to match the models with a similar mass.

A possible explanation for this case includes heating during a CE phase or post-CE phase evolution (see Section \ref{subsec:evolun_path} for further explanation). These processes may have influenced the secondary WD's properties.

\subsection{Periodicity in the light curve}
As outlined in Section \ref{subsec:lc_analysis}, two methods were employed to investigate potential periodicities in the residual light curve that could be indicative of pulsations. Initially, we applied a masking approach exclusively to the eclipses, revealing a distinct periodic variation primarily observed in the SOAR, ULTRACAM $g_s$, and $i_s$ bands. The corresponding FT results are presented in the top three panels of Figure \ref{LC_pulsation}, indicating a peak at a frequency of approximately $\sim$ 197.324594 $\mu Hz$ (equivalent to a period of 1.41 hours). However, this periodicity was not consistently detected in other datasets, such as ULTRACAM $u_s$ and $r_s$, possibly due to higher cadence or increased noise levels.

Upon subtracting the binary model generated using PHOEBE, the same periodicity was not evident in the residual light curve in both cases, whether the Doppler beaming effect was taken into account during the PHOEBE model calculation (Figure \ref{LC_pulsation_masked}) or was not considered (Figure \ref{LC_pulsation_masked_noDoppler}). Nevertheless, this inconsistency raises doubts about the cause of this period.

\begin{figure}
\centering
   \includegraphics[width=8.8cm]{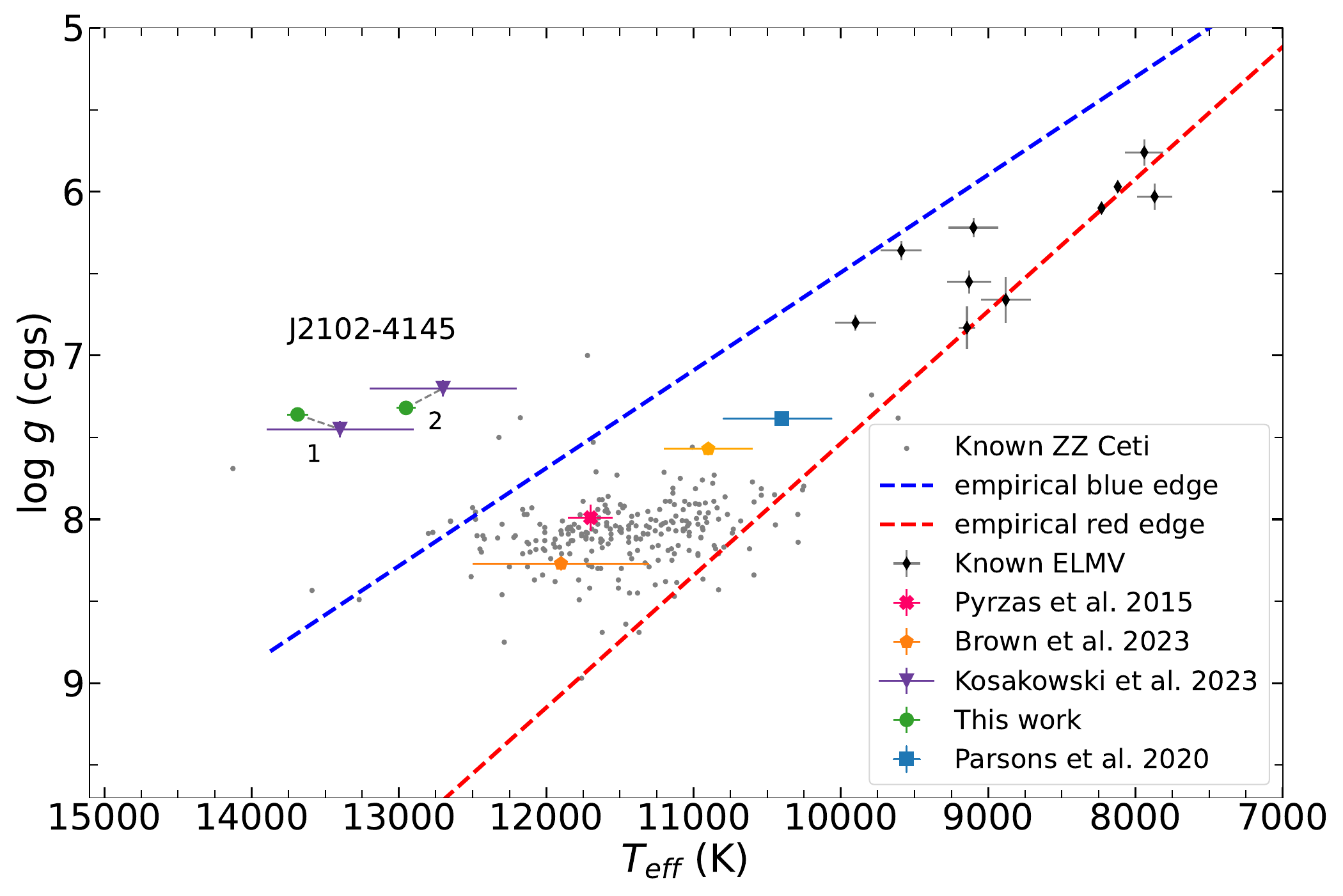}
     \caption{The position of the primary and the secondary stars of the binary system presented in this work, in the $T_{\rm eff}-\log{g}$ plane, are presented as green circles.  The values of $T_{\rm eff}$ and $\log{g}$ found in \citet{2023ELMSouth2} for the same system are depicted as purple triangles. The other two pulsating WD in an eclipsing system found by \citet{2023Brown} are depicted as orange pentagons. The system found by \cite{2015Pyzas} is depicted as a pink $\times$. The first confirmed pulsating WD in a double-degenerate eclipsing system from \citet{2020Parsons} is depicted as a blue square. The known ZZ Ceti \citep{2022MRomero} and ELMVs \citep{2012Hermes, 2013Hermesa, 2013Hermesb, 2015Kilic, 2015Bell, 2017Bell, 2018Pelisolib, 2021Lopez} are shown as grey dots and black diamond shapes, respectively. The ELMV found by \citet{2021Guidry} is not being depicted since its atmospheric parameters have not been determined. The empirical ZZ Ceti instability strip published in \citet{2015Gianninas_ELM6} is marked with dashed lines.}
     \label{teff_logg}
\end{figure}

However, it is worthwhile to note that the values derived in this study for effective temperature and $\log{g}$ placed both the primary and secondary stars outside the ZZ Ceti instability strip, as indicated by the green circles in Figure \ref{teff_logg}. These findings align with the results presented in \citet{2023ELMSouth2} (illustrated by purple triangles in Figure \ref{teff_logg}) for this same system.

In conclusion, the identified periodicity at a period of $P=1.41$ hours in specific light curves introduces intriguing possibilities but remains uncertain in its classification as a pulsation period and its verisimilitude. The potential impact of modelling artefacts and the contextual positioning of the system outside the ZZ Ceti instability strip underscores the need for a cautious interpretation of this observation.


\subsection{Possible evolutionary path}
\label{subsec:evolun_path}

Assuming that the primary (hottest) and most massive WD formed later implies that its progenitor was initially the less massive star in the system. This can be understood if we assume that the system underwent a first phase of mass transfer, which must have been stable, during which the progenitor of the low-mass WD lost its envelope while the mass of its main sequence (MS) companion increased. Given the current short orbital period ($P_{orb}=2.4hr$), the system must have later experienced a common envelope phase \citep[CE,][]{1976Paczynski} when the progenitor of the hotter WD evolved towards the RGB. 

In order to reconstruct the possible evolutionary history of this system, as well as its future, we used the Binary Star Evolution (\bse) code from \citet{hurley2002}.
We started with initial binary systems composed of two MS stars and orbital periods below $\sim$3 days. The short orbital period ensures that the initially more massive star filled its Roche lobe before reaching the RGB phase, where the deep convective envelope would have most likely led to a CE phase. 
For the second mass transfer phase (the CE phase), \bse\, calculates the period after the CE is ejected using the standard $\alpha$-formalism, i.e.
\begin{equation}\label{eq:alpha}
|E_\mathrm{bind}| = \alpha_{CE}\Delta E_\mathrm{orb},
\end{equation}
where $E_\mathrm{bind}$ is the binding energy of the envelope, $\Delta E_\mathrm{orb}$ is the change in orbital energy during the CE phase, and $\alpha_{CE}$ is the CE efficiency, which corresponds to the fraction of $\Delta E_\mathrm{orb}$ that is used to unbind the envelope.
The binding energy is calculated as:
\begin{equation}\label{eq:Ebin}
|E_\mathrm{bin}| = \frac{G M_\mathrm{d} M_\mathrm{d,e}}{\lambda R},
\end{equation}
where $M_\mathrm{d}$, $M_\mathrm{d,e}$ and $R$ are the total mass, envelope mass and radius of the donor at the onset of the CE phase. The binding energy parameter $\lambda$ depends on the structure of the donor star, as well as on the inclusion or not of extra energy sources (in addition to gravitational and orbital) that might assist in unbinding the envelope. Some sources of additional energy suggested in the literature include thermal energy, hydrogen recombination energy, or enthalpy, but their contribution is still on debate \citep[see][for a comprehensive discussion]{Ivanova2013}. Nonetheless, any additional energy included would result in a reduced binding energy, i.e. a larger value of $\lambda$.

The outcome of the CE phase is strongly dependent on the assumed values for $\alpha_{CE}$ and $\lambda$. While $\alpha_{CE}$ must be set as an input parameter in \bse, $\lambda$ can either be set as a fixed value, or one can let the code calculate it. In the latter, one can also decide whether to include a fraction of the recombination energy. Given all the uncertainties in calculating $\lambda$, especially regarding what other potential sources of energy should be included, we have decided to set a fixed value of $\lambda=1$. The CE efficiency $\alpha_{CE}$, on the other hand, was varied to ensure that the simulated systems survived the CE phase, resulting in very close but still detached binary systems.

\begin{figure*}
\centering
   \includegraphics[width=0.85\textwidth]{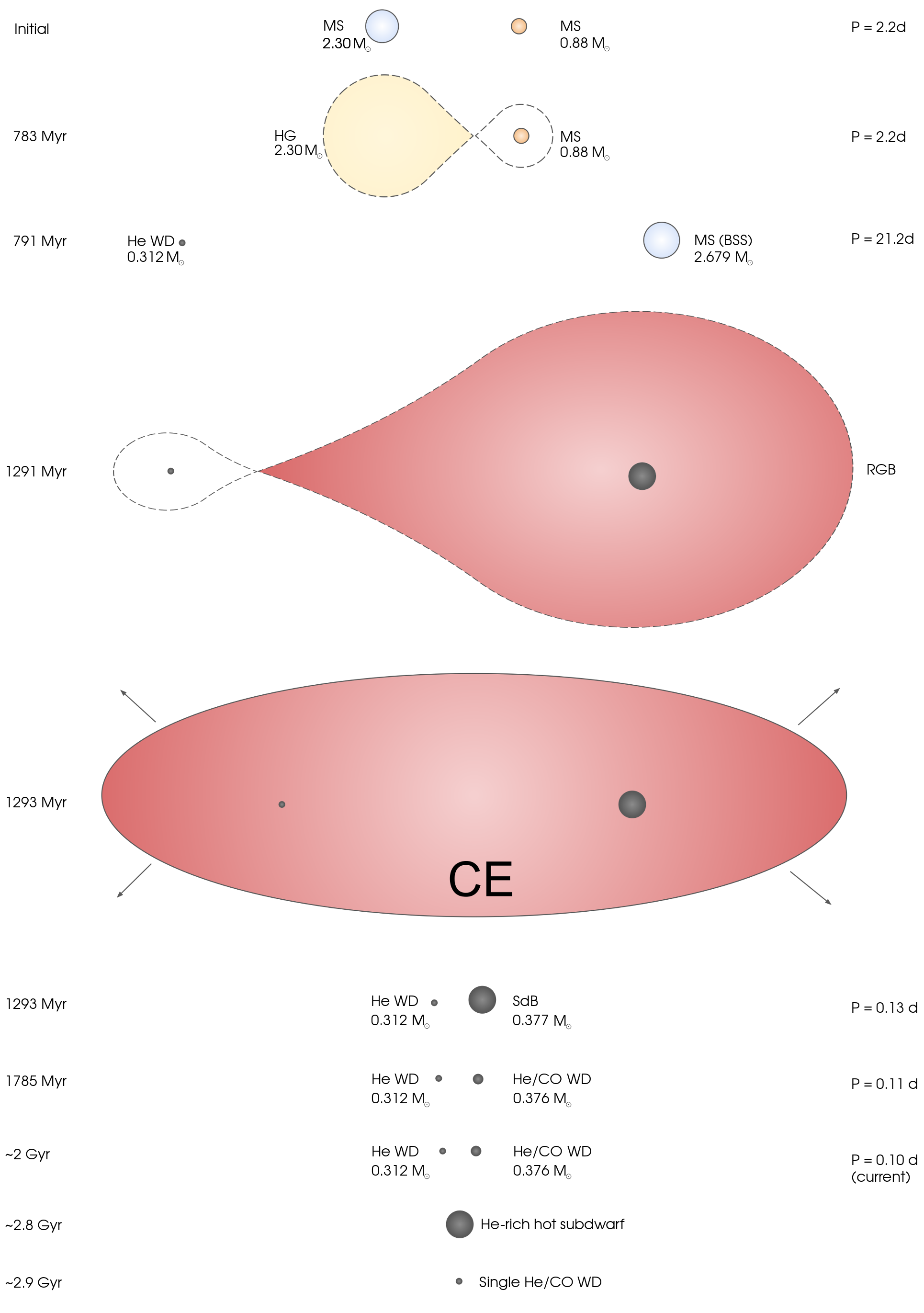}
     \caption{Example of a possible evolutionary path towards the observed system obtained with the \bse\, code from \citet{hurley2002}. The future of the system is also illustrated, based on the current masses and orbital period and assuming that angular momentum loss comes only from gravitational radiation.}
     \label{fig:evolution}
\end{figure*}

We managed to reproduce systems similar to the one observed using initial masses in the range of $\sim2-2.5$\,\Msun\, for the primary and $\lappr1$\,\Msun\, for its companion, with initial orbital periods in the range of $\sim$2-3 days, which caused the more massive star to fill its Roche lobe either towards the end of the MS or on the subgiant branch (Hertzsprung gap). An example of a possible evolutionary path obtained with the \bse\, code is shown in Figure\,\ref{fig:evolution}. The initial system in this sample is composed of two MS stars with masses of $2.30$\,\Msun\, and  $0.88$\,\Msun\, at an orbital period of 2.2 days. The more massive star fills its Roche lobe on the Hertzsprung gap (HG). Initially, mass transfer from the more massive to the less massive star gradually decreases the orbital distance. However, when the donor star has transferred enough mass to its companion, the mass ratio reverses, and the orbital distance begins to increase. The donor star has time to reach the RGB phase, but as the orbital distance increases and the donor has a greatly reduced envelope, mass transfer remains stable. The mass transfer rate gradually decreases until the donor detaches from its Roche lobe and quickly becomes a low-mass helium WD, whose mass is not sufficient to ignite helium. 
The companion star, which accreted much of the mass transferred by the donor, becomes a blue straggler star (BSS), with a much larger mass than initially. This increase in mass accelerates its evolution and, since the orbital period is now longer than initially, it fills its Roche lobe when it is already on the RGB. 

With a much larger mass ratio and the donor now in the RGB phase, this second mass transfer phase becomes dynamically unstable, triggering a CE phase. A very high CE efficiency ($\alpha_{CE}>5$) was needed in order to survive the CE phase without merging. In the particular case we illustrated in Figure\,\ref{fig:evolution} we used $\alpha_{CE}=6.6$, in order to match the observed period at the predicted current time. This was based on the cooling age of the more massive WD in the observed system, which was estimated to be between $\sim200$ and $\sim300$ Myr based on the tracks for low-mass WDs from \citet{2013Althaus}.
For several of the configurations that resulted in simulated systems similar to the observed one, the core of the giant that expelled its envelope during the CE phase had enough mass to ignite helium, becoming a hot subdwarf star (sdB). For a mass of $\sim0.37$\Msun, the stable core helium burning phase lasts about 500 million years \citep{Arancibia23}, until helium is depleted in the core. The star then goes through a brief sdO phase (a few million years, not shown in the figure) and ultimately becomes a WD, likely with a hybrid (He/CO) composition. 
The detached system is now composed of two low-mass WDs at a very short orbital period, which is decreasing due to gravitational radiation, until the two remnants merge to form a single WD. Given the observed orbital period and masses, this should occur in around 800\,Myr (regardless of the previous evolution).
The merger of the two low-mass WDs will trigger helium ignition, leading to the formation of a helium-rich hot subdwarf star undergoing core helium burning for $\sim100$\,Myr, after which the object will evolve into a hybrid He/CO WD \citep[see e.g.,][]{zhangJeffrey2012, Dan2014,Schwab2018}.
However, we decided not to include the mass of the resulting remnant in this figure, because the outcome of the merger also depends on how fast the merger process occurs \citep{zhangJeffrey2012}. 

We note that the example shown here is just an illustrative case of a possible evolutionary path given by the \bse\, code for a similar system, and we are not claiming that this was really the evolution experienced by J2102-4145. Here we discuss some uncertainties with respect to the evolution. 

1) The CE phase is poorly understood, and especially the efficiency parameter ($\alpha_{CE}$) and the structural parameter ($\lambda$) are not well constrained. A low efficiency ($\alpha_{CE}\sim0.3$) has been derived from the observed samples of close WDs with unevolved companions, including low-mass (spectral type M) MS companions \citep{zorotovicetal10-1,Toonen2013,camacho2014}, intermediate-mass (spectral type K to A) MS companions \citep[][and references therein]{hernandez22}, or brown dwarfs \citep{zorotovic2022}. For double WD systems, a larger efficiency ($\alpha_{CE}\sim2$, assuming $\lambda=1$) was favoured by \citep{nelemans2000}, while recently \citet{Scherbak23} also derived a low efficiency ($\alpha_{CE}\sim1/3$) when backtracking the second mass transfer phase in double degenerate systems with low-mass (helium-core) WDs. Reconstructing the evolution of J2102-4145 with \bse\, requires an unusually high efficiency ($\alpha_{CE}\lambda>5$), which seems to be physically implausible. Nonetheless, a recent study by \citet{renzo23} demonstrated that rejuvenated stars, such as the proposed progenitor for the more massive WD in this system, possess significantly less bound envelopes. This implies that the binding energy parameter $\lambda$ should be considerably larger than that of normal stars, resulting in a lower value for $\alpha_{CE}$, which might be consistent with $\alpha_{CE}<1$. It is worth noting, however, that the calculations in \citet{renzo23} focused on massive rejuvenated stars, that will evolve into either a neutron star or a black hole. The authors suggest that a similar scenario may apply to lower-mass accretors, as far as they have a convective core on the MS ($M_{ZAMS}\,\gappr\,1.2$\,\Msun\,). In the example we present, the accretor has a smaller initial mass, so it likely had a radiative core and a convective envelope during the MS phase, before accreting the mass from its initially more massive companion. Consequently, the impact of accretion on the envelope's binding energy remains uncertain in this specific case.

2) Depending on the mass transfer rate, stable mass transfer can either be near fully conservative or highly non-conservative. By examining the total mass before and after the stable mass transfer phase in Figure\,\ref{fig:evolution}, we find that the first mass transfer phase was highly conservative, with a total mass loss from the system of approximately 6 per cent. The \bse\, code does not provide the option to modify this conservativeness. However, a less conservative stable mass transfer phase would have resulted in a larger orbital separation at the end of mass transfer and a lower mass for the progenitor of the more massive WD. The increased orbital separation would have allowed the companion star to fill its Roche lobe later on the RGB, attaining a core mass similar to the observed one when its radius was larger. With a less massive and more extended envelope, the process of envelope ejection would have been facilitated, resulting in a reduced value of $\alpha_{CE}$.

3) The core helium burning phase (as an sdB star) may or may not have occurred, depending on the mass that the progenitor of the secondly formed WD had before losing the envelope. For a core with less than 0.4\Msun\, to ignite helium after losing the envelope, the progenitor must have been more massive than $\sim1.8$\,\Msun\, \citep{han2002,Arancibia23}. Less massive progenitors develop highly degenerate cores during the RGB phase and therefore require a higher core mass to trigger helium burning. Therefore, if the mass transfer was substantially less conservative than in our example, with more than $\sim1$\,\Msun\, lost from the system, a $0.37$\,\Msun\, core would not have ignited helium and would have quickly become a helium WD after the CE phase. To test the effects of mass transfer conservativeness on the evolution, a more flexible and detailed stellar evolutionary code, such as \mesa\,\citep{paxton2011}, should be used. However, this goes beyond the scope of this work, as we are only presenting a possible evolutionary scenario. 
The occurrence of a hot subdwarf phase could provide an explanation for the high effective temperature of the less massive WD. As we saw in Section\,\ref{subsec:results_fit}, the less massive WD in J2102-4145 has a very high effective temperature for the derived mass and radius. It is unlikely that the CE phase is responsible for the higher temperature, because this phase is too short and the accretor is not expected to retain any mass, and therefore any increase in effective temperature should rapidly diminish after the envelope ejection. The sdB phase, on the other hand, can last for approximately half a Gyr for a $0.37$\,\Msun\, sdB, which could be sufficient to produce a long-lasting effect on the temperature of a close companion, especially if there was wind mass transfer from the hot subdwarf (more likely during the sdO phase) to the lower mass WD. 
If the core helium-burning phase is necessary, it implies that the assumption of highly conservative stable mass transfer is not too far from reproducing the actual evolutionary history of the system.

\section{Summary}
In this section, we present the key results and findings from our investigation of the eclipsing double-lined and double-degenerate WD binary system, J2102-4145. This system was previously confirmed in the study by \citet{2023ELMSouth2}, and our research builds upon and enhances our understanding of its properties.

Our study involved an extensive observational campaign, which spanned multiple nights and included data collected from three different telescopes, NTT/ULTRACAM, SOAR/Goodman, and SMARTS-1m, in which we acquired approximately 28 hours of high-speed photometric data. Also, while \citet{2023ELMSouth2} had performed a comprehensive spectral analysis of J2102-4145, our study contributed supplementary spectral analysis conducted independently with two different telescopes: Gemini South/GMOS and VLT/X-Shooter.

Our observations provided crucial insights into the orbital characteristics of J2102-4145. This included the determination of parameters such as the orbital inclination of $i=88.693\substack{+0.006\\ -0.005}$\,deg and the orbital period of $P_{orb}=2.4$hr, in which the last had been previously known from the work of \citet{2023ELMSouth2}.
A central focus of our study was to achieve highly precise parameter measurements for the \target\, binary system. We placed particular emphasis on accurately determining the masses of both WDs within the binary, giving us the values of $M_1=0.375 \pm0.003\,\text{M}_\odot$ for the primary and $M_2=0.314\pm0.003$\,\text{M}$_\odot$ for the secondary. 
In addition to mass measurements, our observations allowed us to determine the radii, $T_{\text{eff}}$ and $\log g$ of both WD components: $R_1=0.0211\pm0.0002$\,R$_\odot$, $T_1=13\,688^{+65}_{-72} $\,K and $\log g_1=7.36\pm0.01$\,dex for the primary and $R_2=0.0203^{+0.0002}_{-0.0003}$\,R$_\odot$, $T_2=12\,952^{+53}_{-66}$\,K and $\log g_2 = 7.32\pm0.01$\,dex for the secondary. This radius information adds to the overall understanding of the system's properties.

From the X-Shooter spectrum, we obtained radial velocity measurements that included semi-amplitude values of $K_1 = 220.8\pm0.7$ km/s for the primary star and $K_2 = 184.6\pm0.8$ km/s for the secondary star. Similar values were also found by \citet{2023ELMSouth2}. The consistency between our results and those of previous studies significantly supports our confidence in the conclusions drawn from this analysis.

Also, our investigation into potential periodicities in the residual light curve, employing two methods, revealed a periodic variation of approximately 1.41 hours in specific bands. However, the inconsistent detection of this periodicity across all datasets raises doubts about its classification as a pulsation. Additionally, our analysis placing both stars outside the ZZ Ceti instability strip aligns with previous findings. The cause of the identified periodicity, while intriguing, remains uncertain in its nature and authenticity as a pulsation period. Considering the potential impact of modelling artefacts and the system's contextual position outside the ZZ Ceti instability strip, caution is warranted in interpreting this observation.

An intriguing finding was the temperature and radius discrepancy observed in the less massive WD. The secondary star appeared hotter than expected for its mass and radius, prompting discussions about possible explanations, such as heating during a CE phase or post-CE phase evolution.
We discussed a possible evolutionary path for the \target\, system, which included initial mass transfer, a CE phase, and the eventual formation of two low-mass WD. While this serves as an illustrative example, it highlights the complexity of the system's evolutionary history.
Acknowledging uncertainties, particularly concerning the CE phase and mass transfer conservativeness, we considered the occurrence of a hot subdwarf phase as a potential explanation for the high $T_{\text{eff}}$ of the less massive WD.

In conclusion, our study contributes valuable insights into the \target\, system, shedding light on its fundamental properties and potential evolutionary history. These findings enrich our understanding of binary WD systems and the intricate pathways they traverse.

\begin{acknowledgements}
We thank the anonymous referee for their suggestions, which helped improve our manuscript. LAA acknowledges financial support from CONICYT Doctorado Nacional in the form of grant number No: 21201762 and ESO studentship program.
Based on observations collected with the GMOS spectrograph on the 8.1~m Gemini-South telescope at Cerro Pachón, Chile, under the program GS-2022A-Q-416, and observation with Goodman SOAR 4.1~m telescope, at Cerro Pachón, Chile, and SMARTS-1m telescope at Cerro Tololo, Chile, under the program allocated by the Chilean Time Allocation Committee (CNTAC), no: CN2022A-412490, and SOAR observational time through NOAO program SO2021A-008\_0516. JM was supported by funding from a Science and Technology Facilities Council (STFC) studentship. ULTRACAM operations and VSD are funded by the Science and Technology Facilities Council (grant ST/V000853/1).
M.V. acknowledges support from FONDECYT (grant No: 1211941). 
I.P. acknowledges a Warwick Astrophysics prize post-doctoral fellowship made possible thanks to a generous philanthropic donation.
P.N. acknowledges support from the Grant Agency of the Czech Republic (GA\v{C}R 22-34467S).
The Astronomical Institute in Ond\v{r}ejov is supported by the project RVO:67985815.
This research has used the services of \mbox{\url{www.Astroserver.org} under reference XG61YQ}.
\end{acknowledgements}

%
%

\bibliographystyle{aa} 
\bibliography{example} 



\appendix

\section{Extra GMOS/Gemini-South spectra}
\label{appendix:gmos_extra}
\begin{figure*}[h!]
   \centering
   \includegraphics[width=0.9\textwidth]{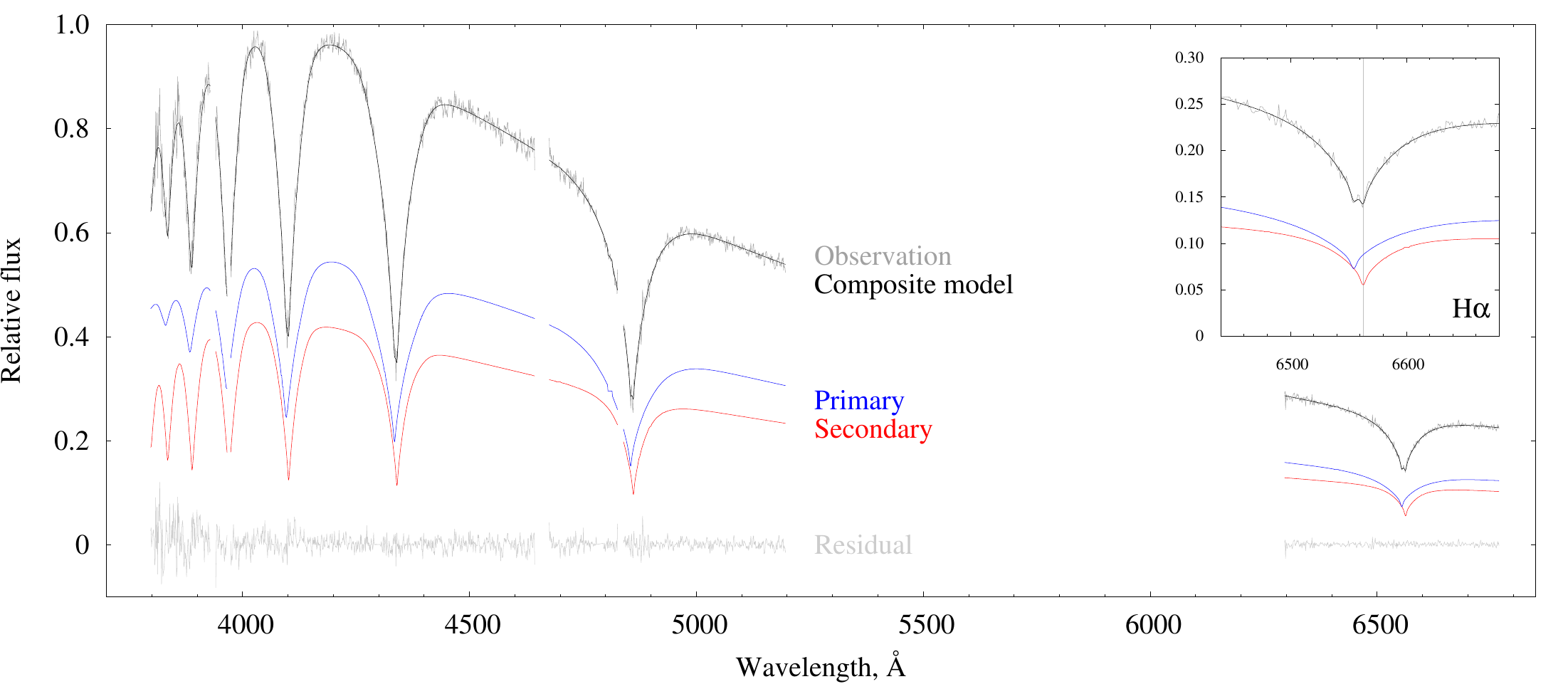}
      \caption{
      Best-fit {\sc XT{\fontsize{6}{8}\selectfont
 GRID}}/WD model to the Gemini/GMOS spectrum of \target\ at orbital phase $\phi=0.75$ (maximum radial velocity difference) in the rest frame of the secondary. 
      The observed spectrum can be reproduced by two nearly identical WD components ({\sl red and blue lines}). 
      The phase-resolved spectral coverage for the entire orbit is available in Figure\,\ref{H_alpha_beta_GMOS}. The wavelength gap between $\sim$520--625\,nm\ is due to the bad amplifier number 5 (see Section \ref{SUBSection_spectroscopy}). 
      { The flux of the observation was adjusted to the flux level of the theoretical composite continuum.}
      }
\label{SpFit_gemini}
\end{figure*}

\begin{figure*}[h!]
\centering
   \includegraphics[width=0.9\textwidth]{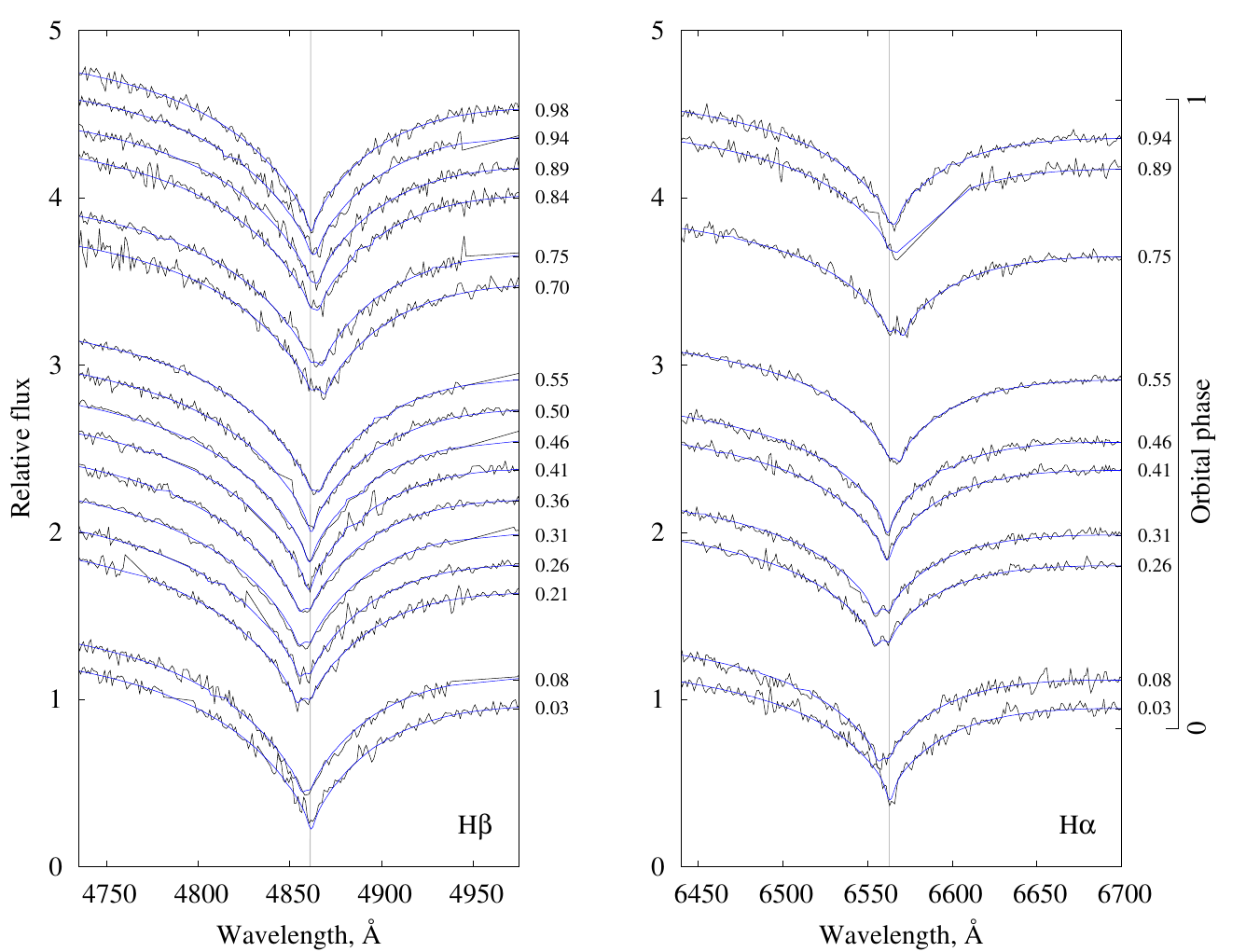}
     \caption{
Best-fit {\sc XT{\fontsize{6}{8}\selectfont
 GRID}}/WD models and the spectral evolution of the composite H$\alpha$ (right panel) and H$\beta$ (left panel) composite line profiles in the Gemini/GMOS observations in the rest frame of the primary. The fluxes of the observations were adjusted to the flux level of the theoretical composite continuum.
}
     \label{H_alpha_beta_GMOS}
\end{figure*}

\section{Light curve analysis for periodicity}
\label{appendix:lc_periodicity}

\begin{figure*}[h!]
\centering
   \includegraphics[width=\textwidth]{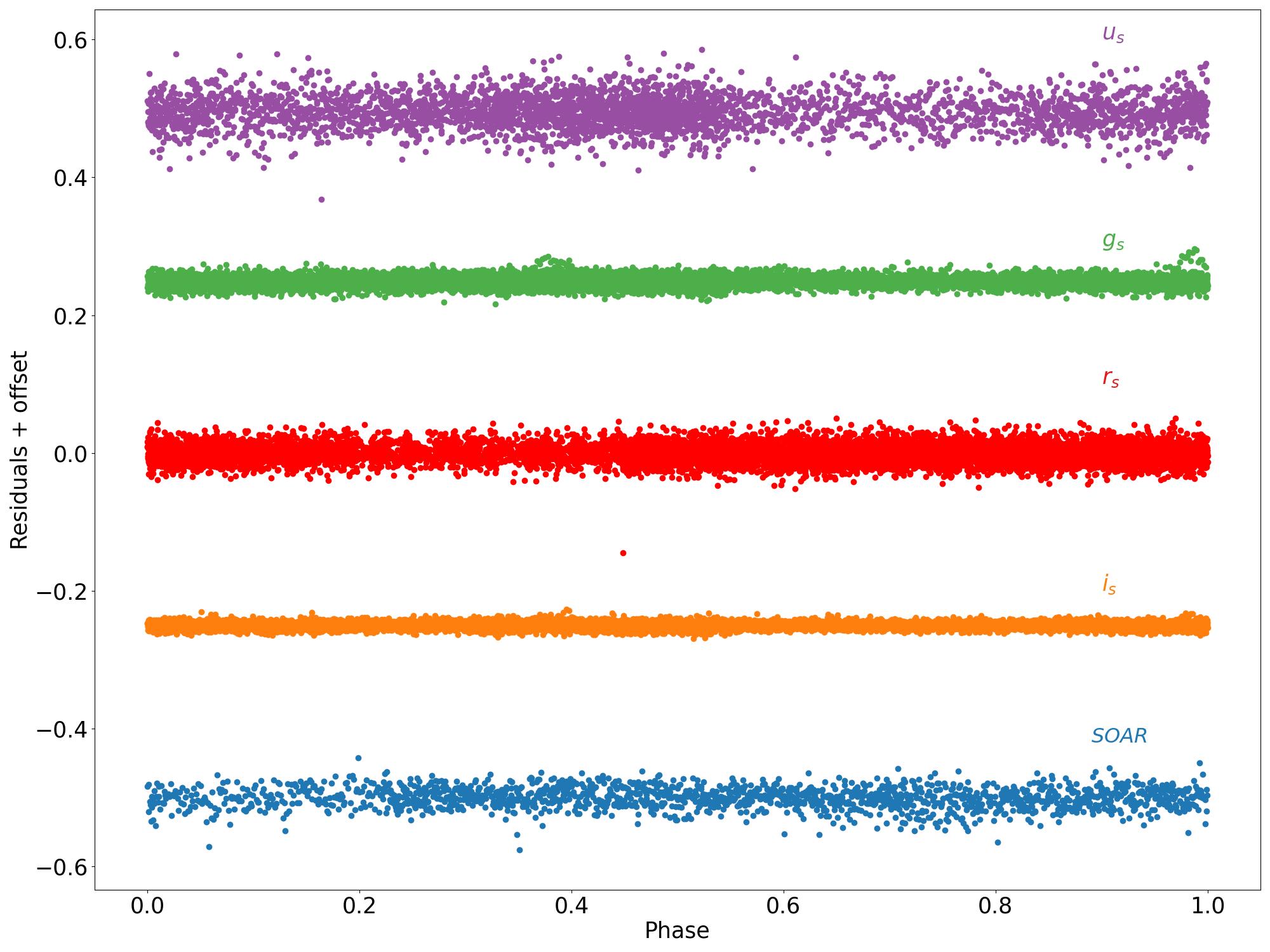}
     \caption{
     ULTRACAM light curve residuals after subtraction of the binary model in $u_s$, $g_s$, $r_s$, $i_s$ and SOAR, respectively.
     }
     \label{lc_residuals}
\end{figure*}


 \begin{figure*}
\centering
   \includegraphics[width=17cm]{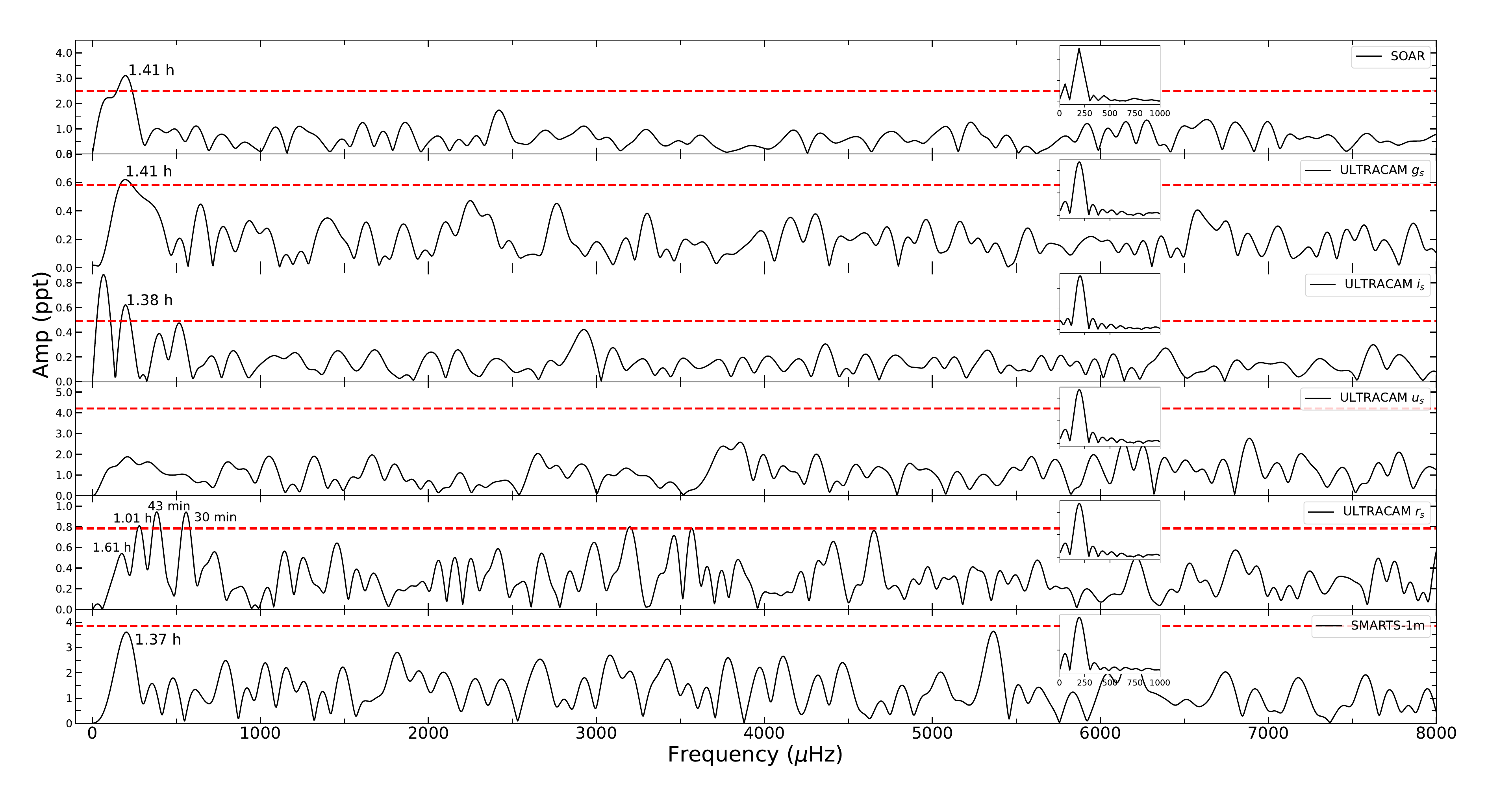}
     \caption{FT for \target\, for SOAR, ULTRACAM $u_{s}$, $g_{s}$, $i_{s}$, $r_{s}$ and SMARTS-1m observations. The 1/1000 False Alarm Probability detection limit (red dashed line) was computed using random shuffling of the data. The spectral window for each case is depicted as an inset plot, with the x-axis in $\mu Hz$ and all inset plots being in the same scale. For the $r_{s}$ ULTRACAM data, the combination $(f_1+f_2)/2$ of the periods of $1.61$~h (171.872 $\mu Hz$) and 1.01~hr (274.345 $\mu Hz$) its half of the orbital period.}
     \label{LC_pulsation}
\end{figure*}

 \begin{figure*}
\centering
   \includegraphics[width=17cm]{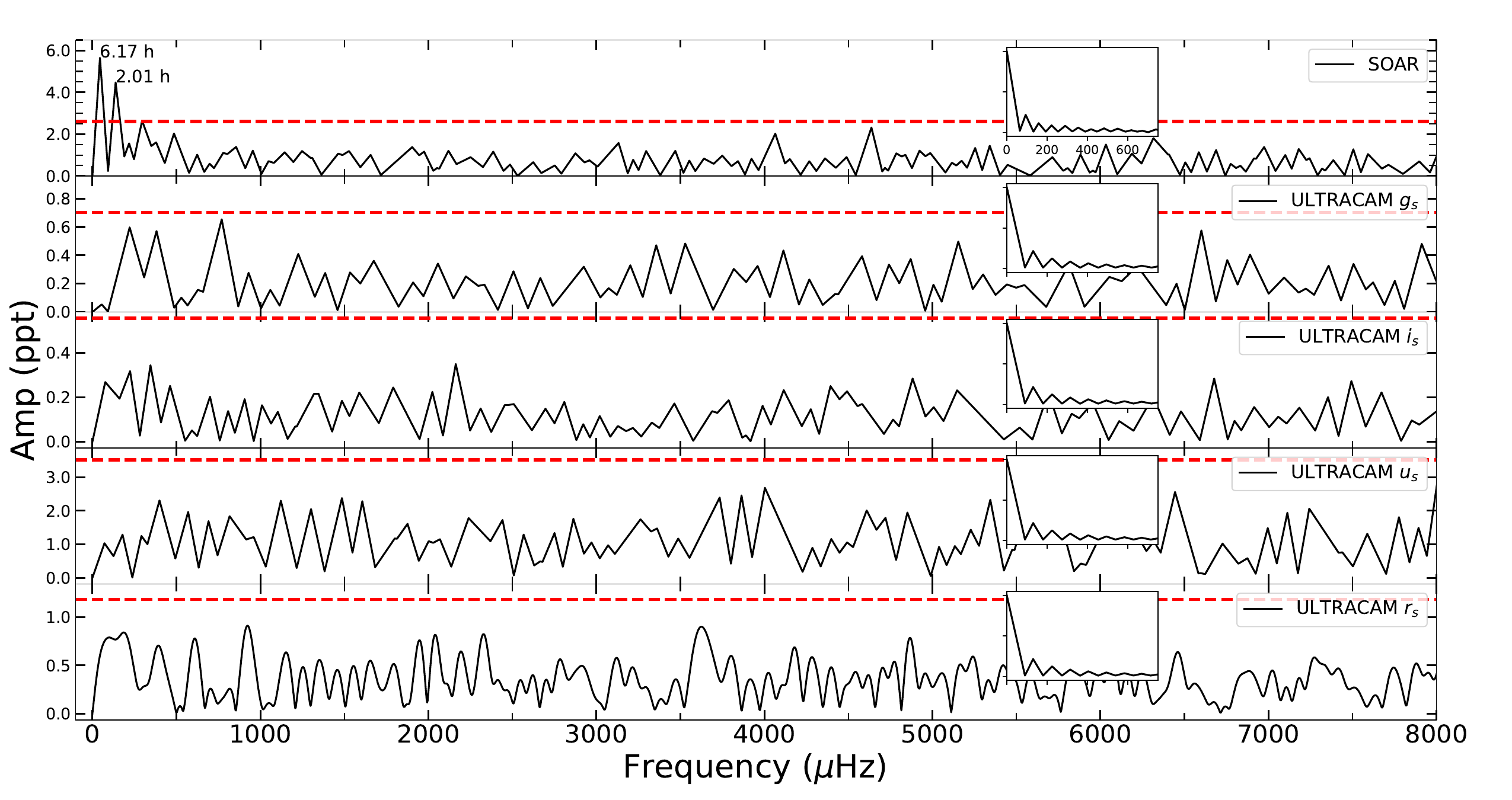}
     \caption{FT for \target\, for SOAR, ULTRACAM $u_{s}$, $g_{s}$, $i_{s}$, $r_{s}$ for the residual light curve subsequent to subtracting the light curve fit generated using PHOEBE (see Section \ref{subsec:lc_analysis} for further information). The 1/1000 False Alarm Probability detection limit (red dashed line) was computed using random shuffling of the data. The spectral window for each case is depicted as an inset plot, with the x-axis in $\mu Hz$ and all inset plots being in the same scale. For the SOAR data, although the 6.17\,hr is above our threshold, it exceeds the duration of our light curve, which makes it less reliable.}
     \label{LC_pulsation_masked}
\end{figure*}

 \begin{figure*}
\centering
   \includegraphics[width=17cm]{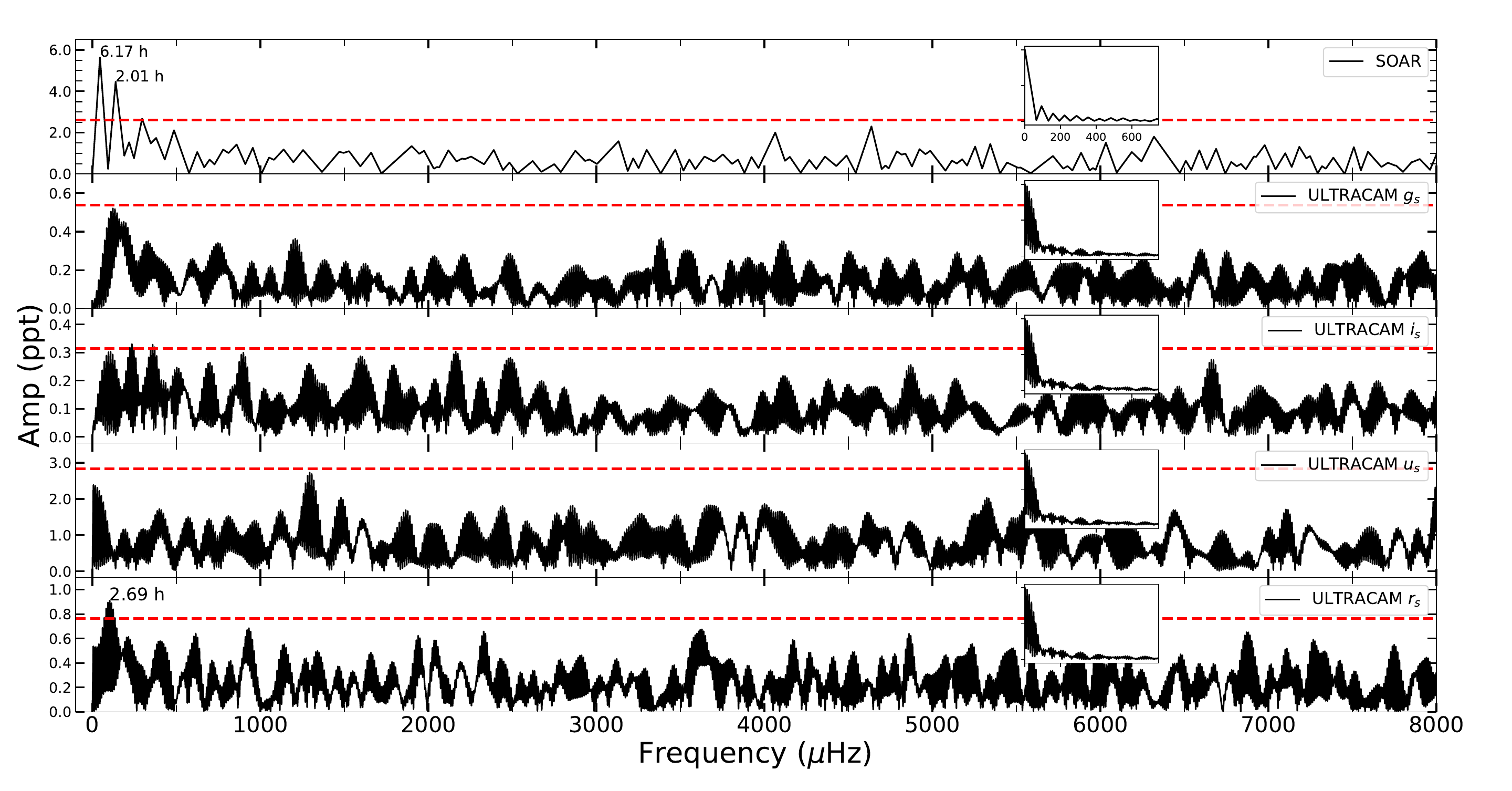}
     \caption{For the target \target\ observed with SOAR and ULTRACAM $u_{s}$, $g_{s}$, $i_{s}$, $r_{s}$ FT analysis on the residual light curve, obtained after subtracting the light curve fit generated using PHOEBE (details in Section \ref{subsec:lc_analysis}). The SOAR and ULTRACAM data in $u_{s}$, $g_{s}$, $i_{s}$, and $r_{s}$ bands were used, mirroring the approach undertaken for Figure \ref{LC_pulsation_masked}. However, in this case, a new PHOEBE model was applied without considering the Doppler beaming effect. This adjustment aimed to investigate and eliminate the possibility that the 1.4-hour signal observed with masked eclipses (see Figure \ref{LC_pulsation}) was caused by the Doppler beaming effect.}
     \label{LC_pulsation_masked_noDoppler}
\end{figure*}

\section{Posterior probability distributions}
\label{appendix:phoebe_corner_plot}
 \begin{figure*}
\centering

   \includegraphics[width=\textwidth]{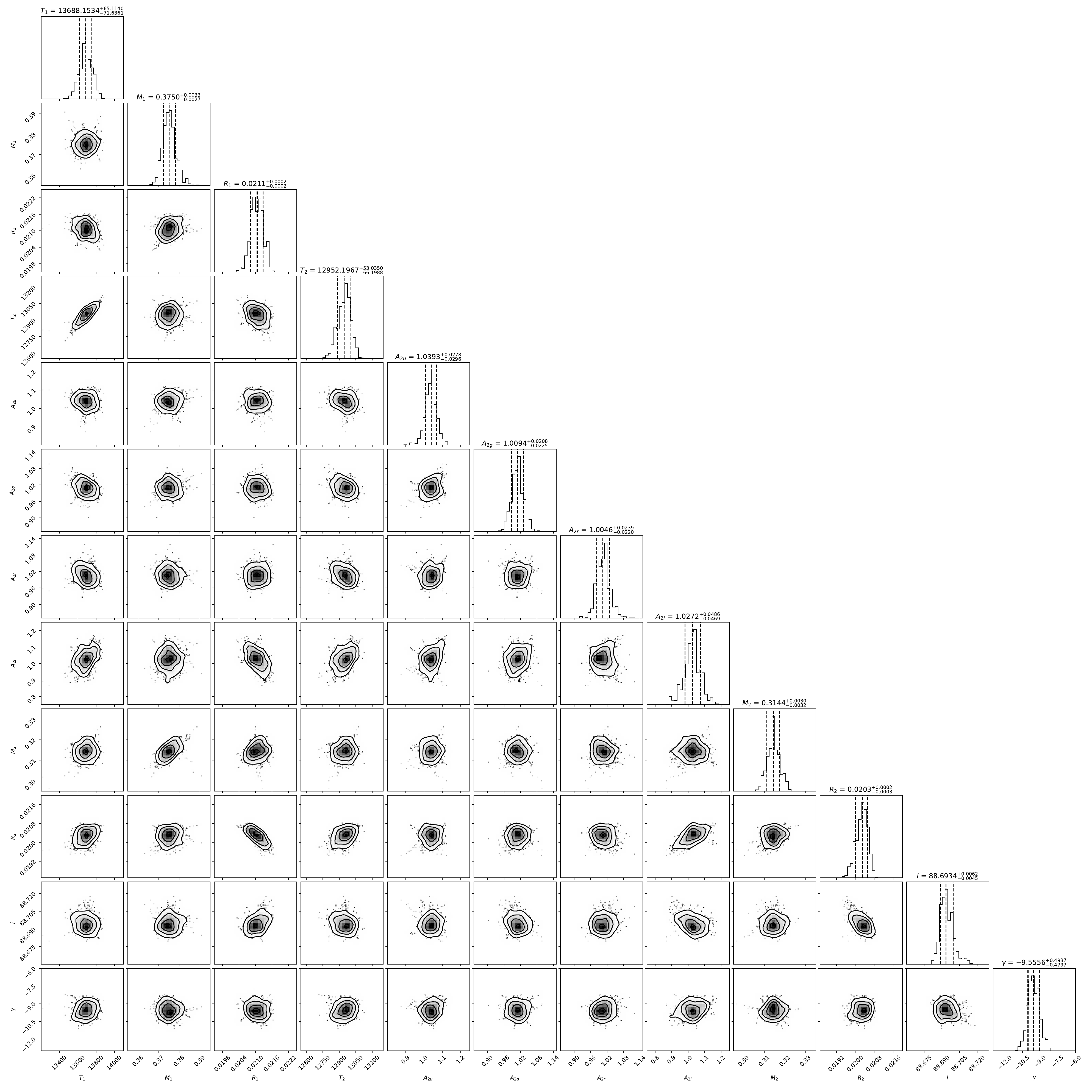}
     \caption{A corner plot of the phoebe2 MCMC (\citealt{2016ForemanMCMC}) posteriors for the 
     primary mass ($M_1$) and radius ($R_1$), the secondary mass ($M_2$), radius ($R_2$) and effective temperature ($T_2$), as well as the binary orbital inclination ($i$)}
     \label{corner_plot}
\end{figure*}

\section{Extra tables}
\begin{table*}[]
    \centering
    \begin{tabular}{c|c|c|c|c}
Phase	&	$RV_1$	(km/s)&	$\sigma RV_1$	&	$RV_2$	(km/s)&	$\sigma RV_2$	\\ \hline
0.013	&	28.43	&	3.2	&	-11.89	&	2.5	\\
0.052	&	83.21	&	2	&	-75.1	&	3	\\
0.059	&	92.41	&	2	&	-71.95	&	4	\\
0.099	&	139.78	&	3.8	&	-114.08	&	8	\\
0.138	&	177.06	&	2	&	-146.5	&	2	\\
0.177	&	204.19	&	2	&	-172.8	&	2.5	\\
0.216	&	218.48	&	3	&	-188.47	&	3	\\
0.255	&	219.72	&	3	&	-187.95	&	4	\\
0.294	&	207.5	&	2	&	-172.61	&	2.3	\\
0.333	&	182.91	&	2	&	-157.21	&	2.3	\\
0.372	&	147.1	&	2.4	&	-127.02	&	3.5	\\
0.411	&	102.72	&	2.5	&	-87.46	&	3	\\
0.45	&	51.74	&	6	&	-36.72	&	6	\\
0.489	&	-2.33	&	2	&	-1.16	&	3	\\
0.529	&	-57.65	&	2	&	39.22	&	2	\\
0.583	&	-125.69	&	2.6	&	96.01	&	2	\\
0.622	&	-166.2	&	2	&	144.34	&	5	\\
0.661	&	-196.62	&	2	&	165.79	&	4	\\
0.7	&	-215.11	&	2	&	178.19	&	2	\\
0.739	&	-220.79	&	2	&	179.07	&	2	\\
0.779	&	-212.26	&	5	&	187.36	&	7.5	\\
0.818	&	-190.98	&	4	&	176.39	&	6	\\
0.857	&	-158.42	&	4	&	138.96	&	7	\\
0.896	&	-115.97	&	3	&	96.4	&	3.5	\\
0.935	&	-66.64	&	4.3	&	67.07	&	3.3	\\
0.974	&	-13.2	&	4	&	2.48	&	2.5	\\ \hline

\end{tabular}
    \caption{RV values for the primary and secondary star, with its respective uncertainties, as used in Figure \ref{rv_xshooter}.}
    \label{tab:RV_values}
\end{table*}

\begin{table*}[]
    \centering
    \begin{tabular}{c|c|c|c|c|c}
\multicolumn{3}{c|}{ Primary}		& 		\multicolumn{3}{c}{ Secondary} 		\\\hline
MBJD 	&	$\Delta$MBJD	&	Filter	&	      MBJD 	&	$\Delta$MBJD	&	Filter		\\
59410.3609778	&	0.0000085 	&	$g_s$	&	59410.4110843	&	0.0000098 	&	$g_s$		\\
59410.360966	&	0.0000160 	&	$i_s$	&	59410.4110838	&	0.0000129 	&	$i_s$		\\
59411.3630806	&	0.0000086 	&	$g_s$	&	59411.3129575	&	0.0000096 	&	$g_s$		\\
59411.3630658	&	0.0000127 	&	$i_s$	&	59411.3129374	&	0.0000174 	&	$i_s$		\\
59412.0645211	&	0.0000101 	&	$g_s$	&	59411.4131806	&	0.0000123 	&	$g_s$		\\
59412.0645158	&	0.0000139 	&	$i_s$	&	59411.4131637	&	0.0000156 	&	$i_s$		\\
59412.1647239	&	0.0000118 	&	$g_s$	&	59412.1146357	&	0.0000120 	&	$g_s$		\\
59412.1647242	&	0.0000160 	&	$i_s$	&	59412.1146165	&	0.0000177 	&	$i_s$		\\
59737.3421229	&	0.0000101 	&	$g_s$	&	59412.3150527	&	0.0000093 	&	$g_s$		\\
59737.3421435	&	0.0000143 	&	$i_s$	&	59412.3150254	&	0.0000173 	&	$i_s$		\\
59841.0581975	&	0.0000075 	&	$g_s$	&	59412.4152568	&	0.0000073 	&	$g_s$		\\
59841.0582062	&	0.0000129 	&	$r_s$	&	59412.4152507	&	0.0000158 	&	$i_s$		\\
59841.1584092	&	0.0000114 	&	$g_s$	&	59737.3922405	&	0.0000083 	&	$g_s$		\\
59841.1584007	&	0.0000122 	&	$r_s$	&	59737.3922115	&	0.0000231 	&	$i_s$		\\
59842.1604868	&	0.0000087 	&	$g_s$	&	59841.1083025	&	0.0000096 	&	$g_s$		\\
59842.1604901	&	0.0000218 	&	$r_s$	&	59841.1082899	&	0.0000139 	&	$r_s$		\\
59842.2606987	&	0.0000108 	&	$g_s$	&	59842.1103795	&	0.0000114 	&	$g_s$		\\
59842.2607014	&	0.0000129 	&	$r_s$	&	59842.1103702	&	0.0000140 	&	$r_s$		\\
	&		&		&	59842.2105978	&	0.0000121 	&	$g_s$		\\
	&		&		&	59842.2106085	&	0.0000168 	&	$r_s$		\\ \hline

\end{tabular}
    \caption{A table of centre of eclipse times. The first set of timings are for the primary eclipse and those of the secondary are presented after.}
    \label{tab:my_label}
\end{table*}

\end{document}